\documentclass[a4paper,11pt]{article}
\pdfoutput=1 

\usepackage{jheppub} 

\usepackage[T1]{fontenc}
\usepackage{multirow,bbold,slashed}

\allowdisplaybreaks
 
\definecolor{nicered}{rgb}{0.7,0.1,0.1} 
\definecolor{nicegreen}{rgb}{0.1,0.5,0.1}
\definecolor{niceblue}{rgb}{0.0,0.1,0.7}
\hypersetup{colorlinks,citecolor=niceblue,linkcolor=niceblue,urlcolor=niceblue}

\usepackage[normalem]{ulem}

\def \beq{\begin{equation}}
\def \eeq{\end{equation}}
\def \bea{\begin{eqnarray}}
\def \eea{\end{eqnarray}}

\title{Higgs production from anomalous gluon dynamics}

\author[a]{Ulrich Haisch}

\affiliation[a]{Max Planck Institute for Physics, \\ Boltzmannstr.~8, 85748 Garching, Germany}

\emailAdd{haisch@mpp.mpg.de} 

\preprint{MPP-2025-38} 

\abstract{We present a two-loop analysis of the contributions to Higgs production via gluon-gluon fusion arising from the triple-gluon operator in the Standard Model effective field theory (SMEFT). Our discussion covers all aspects of renormalization group~(RG) improved perturbation theory, including matching and running within the SMEFT. This~study can therefore be seen as a blueprint of the intricacies and subtleties that arise in RG improved SMEFT calculations for collider processes beyond the leading order.}

\begin{document} 
\maketitle
\flushbottom

\section{Introduction} 
\label{sec:introduction}

The year 2025 marked the 40th anniversary of the effective Lagrangian for new interactions and flavor conservation~\cite{Buchmuller:1985jz}, which is now known as the Standard Model effective field theory~(SMEFT)~\cite{Grzadkowski:2010es, Brivio:2017vri,Isidori:2023pyp}. Since its introduction, the SMEFT has made significant strides and has become a well-established framework for constraining indirect signs of beyond the Standard Model~(BSM) physics at the Large Hadron Collider~(LHC). This effort places rigorous demands on the theoretical Standard Model~(SM) predictions and increasingly also on the precision of the BSM calculations.

While one-loop QCD matching calculations in the SMEFT for LHC processes have seen a degree of automation~\cite{Degrande:2020evl}, and the one-loop renormalization group (RG) evolution of the Wilson coefficients for dimension-six SMEFT operators is a solved problem~\cite{Jenkins:2013zja,Jenkins:2013wua,Alonso:2013hga}, two-loop calculations within the SMEFT have only emerged in recent years. Recent progress includes, but is not limited to, the calculations and studies presented in~\cite{Haisch:2022nwz,DiNoi:2023ygk,Gauld:2023gtb,Heinrich:2023rsd,DiNoi:2024ajj,Born:2024mgz,Haisch:2024wnw,Bonetti:2025hnb}. These works share the common feature of achieving two-loop accuracy either in the matching or the running within the SMEFT, but not both. This article seeks to illustrate the full complexity of RG improved perturbation theory beyond the one-loop level in the SMEFT through a nontrivial yet instructive example. The example we consider is the contribution of the triple-gluon operator to Higgs production in gluon-gluon fusion~($gg \to h$). Assuming that this operator is the dominant ultraviolet~(UV) deformation, the $gg \to h$ production cross section is modified first at the two-loop level. These corrections stem from three distinct types of contributions, which involve matching and running effects at different perturbative orders. The unphysical dependence on the renormalization scale cancels order by order in perturbation theory only when these corrections are properly combined. By~using the relevant SMEFT RG equations~(RGEs), one can then resum the large logarithms that appear in the $gg \to h$ prediction, going beyond leading logarithmic~(LL) accuracy.

Another reason for considering the triple-gluon operator contribution to $gg \to h$ production is that its Wilson coefficient remains relatively weakly constrained by jet observables and top-quark processes~\cite{Ghosh:2014wxa,Krauss:2016ely,Hirschi:2018etq,Goldouzian:2020wdq,Bardhan:2020vcl,Ellis:2020unq}. In fact, the nominally strongest constraints come from multijet production and rely on the strong enhancement of the quadratic contributions in the Wilson coefficient within the high-energy tails of multijet distributions. This raises questions about the reliability of the effective field theory~(EFT) expansion when extracting bounds on the Wilson coefficient of the triple-gluon operator from multijet observables. Given these limitations, indirect precision tests of the triple-gluon operator are of particular interest. In~this work, we show that precise LHC measurements of the $gg \to h$ production cross section can serve as complementary probes of anomalous gluon~dynamics.

This article is organized as follows: in~Section~\ref{sec:framework}, we present the subset of dimension-six operators relevant to our study. Section~\ref{sec:calculation} provides a brief overview of the key steps in computing the SMEFT corrections to $gg \to h$. The relevant beta functions are discussed in~Section~\ref{sec:betas}, while~Section~\ref{sec:formfactor} dissects the structure of the SMEFT corrections to the $gg \to h$ form factor. Section~\ref{sec:RG} explores the solutions of the RGEs. In~Section~\ref{sec:pheno}, we analyze the phenomenological implications of our findings. Our conclusions are presented in~Section~\ref{sec:conclusions}. Additional technical details are provided in~Appendices~\ref{app:realradiation} to~\ref{app:RG}.

\section{Framework} 
\label{sec:framework}

To establish our notation and conventions, we begin by defining the SMEFT Lagrangian:
\beq \label{eq:LSMEFT}
{\cal L}_{\rm SMEFT} = \sum_i C_i (\mu) \hspace{0.25mm} Q_i \,.
\eeq
Here, $C_i (\mu)$ denotes the dimensionful Wilson coefficients evaluated at the renormalization scale $\mu$, which are associated with the corresponding effective operators~$Q_i$. For the entirety of this article, we assume that all Wilson coefficients are real.

In the Warsaw basis of operators~\cite{Grzadkowski:2010es}, the full set of dimension-six operators relevant to this study includes
\beq \label{eq:operators}
Q_{G} = f^{abc} \hspace{0.25mm} G_{\mu}^{a, \nu} G_{\nu}^{b, \rho} G_{\rho}^{c, \mu} \,, \qquad 
Q_{tG} = \bar q \hspace{0.25mm} \sigma^{\mu \nu} \widetilde H \hspace{0.25mm} T^a t \hspace{0.5mm} G_{\mu \nu}^a \,, \qquad 
Q_{HG} = H^\dagger H \hspace{0.25mm} G_{\mu \nu}^a G^{a, \mu \nu} \,.
\eeq
In this context, $G_{\mu \nu}^a$ is the field strength tensor of $SU(3)_C$, with $T^a =\lambda^a/2$ the generators of the group, where $\lambda^a$ are the Gell-Mann matrices and $f^{abc}$ are the structure constants.
The~corresponding gauge coupling will be denoted by $g_s$. $H$ represents the SM Higgs doublet, and we used $\widetilde H = \varepsilon \cdot H^\ast$, where $\varepsilon = i \sigma^2$ is the antisymmetric~$SU(2)$ tensor. The symbol $q$ represents the left-handed third-generation quark $SU(2)_L$ doublet, while~$t$~denotes the right-handed top-quark $SU(2)_L$~singlet. Lastly, notice that for the operator~$Q_{tG}$ the addition of the hermitian conjugate is implied in~(\ref{eq:LSMEFT}). 

Below, we concentrate on a particular class of BSM scenarios in which, at the high-energy scale $\Lambda$, the initial conditions of the Wilson coefficients follow the hierarchy:
\beq \label{eq:hierarchy}
C_{G} (\Lambda) \gg C_{tG} (\Lambda) \gg C_{HG} (\Lambda) \simeq 0 \,.
\eeq
We remain agnostic about how such scenarios are realized, as the primary goal of this article is to elucidate the rich structure of RG improved perturbation theory in the SMEFT. As we will see below, obtaining the full leading SMEFT correction to $gg \to h$ in BSM scenarios of the form~(\ref{eq:hierarchy}) requires accounting for three types of contributions: $(i)$~two-loop matching contributions involving the tree-level Wilson coefficient $C_{G}$, $(ii)$~one-loop matching contributions involving the one-loop evolved Wilson~coefficient $C_{tG}$, and $(iii)$~tree-level matching contributions involving the two-loop evolved Wilson coefficient~$C_{HG}$. 

\section{Calculation} 
\label{sec:calculation}

\begin{figure}[t!]
\begin{center}
\includegraphics[width=0.8\textwidth]{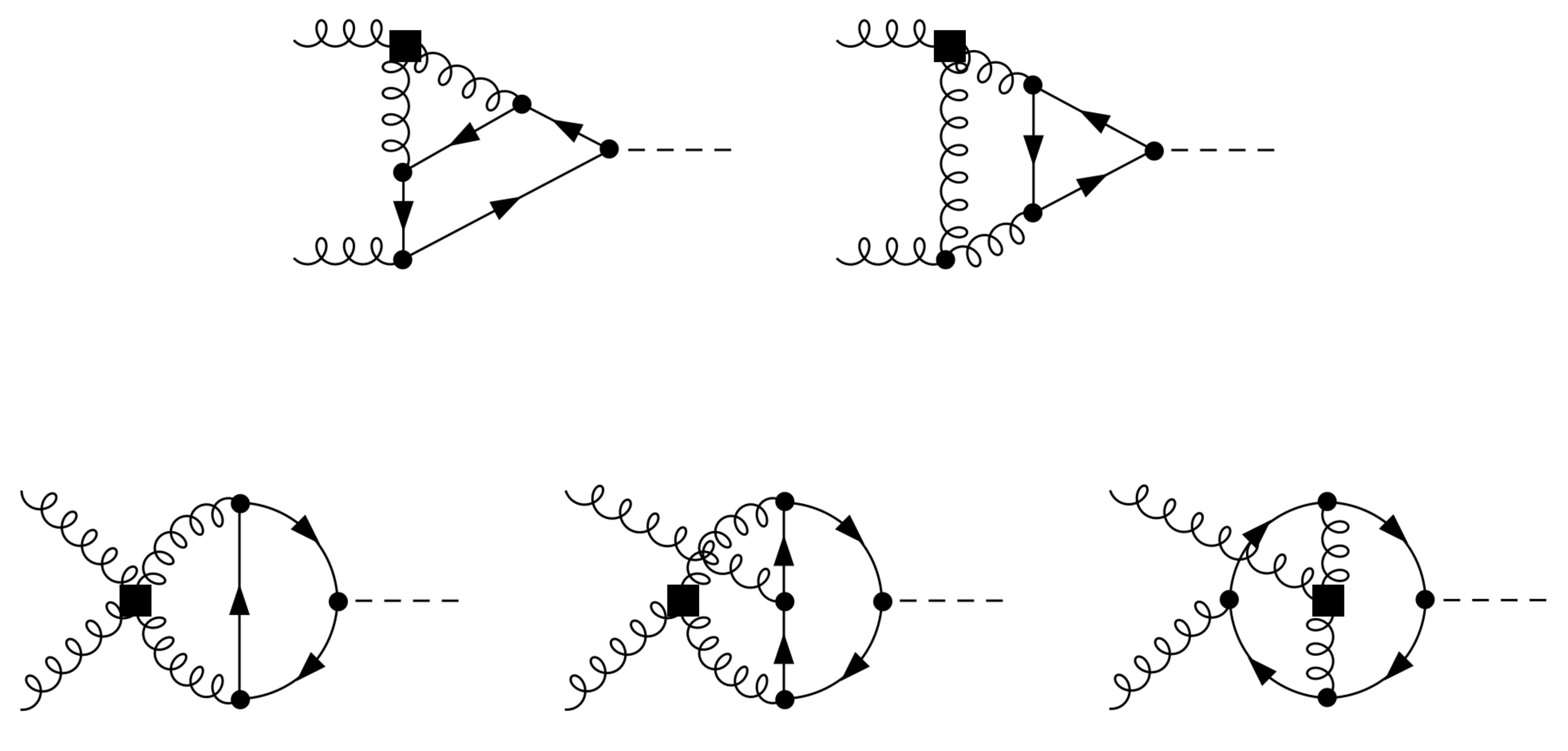}
\end{center}
\vspace{0mm} 
\caption{\label{fig:diagrams1} Examples of two-loop contributions to the $gg \to h$ process involving an insertion of the operator $Q_{G}$ defined in~(\ref{eq:operators}). The operator insertions are represented by black boxes, gluons and the Higgs are depicted by wiggly lines and dashed lines, respectively, while the solid lines depict quarks. See the main text for further details.} 
\end{figure}

We use dimensional regularization in $d= 4 - 2 \epsilon$ to regulate UV and infrared~(IR) singularities, with renormalization scale $\mu$ and a naive anticommuting~$\gamma_5$~(NDR)~\cite{Chanowitz:1979zu}. As no traces involving $\gamma_5$ appear, the NDR scheme is evidently consistent. Our computation used the {\tt Mathematica} packages {\tt FeynRules}~\cite{Alloul:2013bka}, {\tt FeynArts}~\cite{Hahn:2000kx}, {\tt FormCalc}~\cite{Hahn:2016ebn}, and {\tt LiteRed}~\cite{Lee:2013mka}. {\tt FeynRules} implements the SMEFT operators~(\ref{eq:operators}) and generates a {\tt FeynArts} model file, which is used to construct the Feynman diagrams and amplitudes. {\tt FormCalc} projects onto form factors and handles color and Dirac algebra, while {\tt LiteRed} reduces scalar integrals to one-loop and two-loop master integrals~(MIs). All MIs but one are known analytically from~\cite{Anastasiou:2006hc}, and the remaining one is computed using differential equations~\cite{Kotikov:1990kg,Remiddi:1997ny,Gehrmann:1999as,Argeri:2007upc,Henn:2014qga}. All MIs are expressed as Laurent series in $\epsilon$ using harmonic polylogarithms (HPLs). As a cross-check, the MIs were numerically evaluated using~{\tt AMFlow}~\cite{Liu:2022chg}, and the results agreed with the analytical expressions.

Figure~\ref{fig:diagrams1} shows two-loop Feynman diagrams with a single insertion of the $Q_G$ operator~(\ref{eq:operators}) contributing to $gg \to h$. The bare amplitude contains both $1/\epsilon^2$ and~$1/\epsilon$ poles, which are of UV origin and have no IR contributions. This point is explained in Appendix~\ref{app:realradiation}, which covers the real radiation contributions from $Q_G$. To achieve a UV-finite result, one must account for the fact that~$Q_{G}$ mixes into the chromomagnetic top-quark dipole operator~$Q_{tG}$ at one loop, which in turn~mixes into the Higgs-gluon operator~$Q_{HG}$ at the same order. All~relevant one-loop anomalous dimensions can be found in~\cite{Jenkins:2013wua,Alonso:2013hga}. After subtracting the appropriate counterterm contributions in the modified minimal subtraction~($\overline{\rm MS}$) scheme~\cite{Bardeen:1978yd}, the remaining $1/\epsilon$ pole becomes local (for~general discussions of this point see,~e.g.,~\cite{Chetyrkin:1997fm,Gambino:2003zm}). The~left-over UV divergence determines the two-loop anomalous dimensions that govern the mixing of $Q_{G}$ into $Q_{HG}$. Observe that, numerically, only top-quark loops are important due to the strong Yukawa suppression of all other quarks. Hence, the results below include only contributions involving the top-quark Yukawa coupling $y_t$. 

\begin{figure}[t!]
\begin{center}
\includegraphics[width=0.85\textwidth]{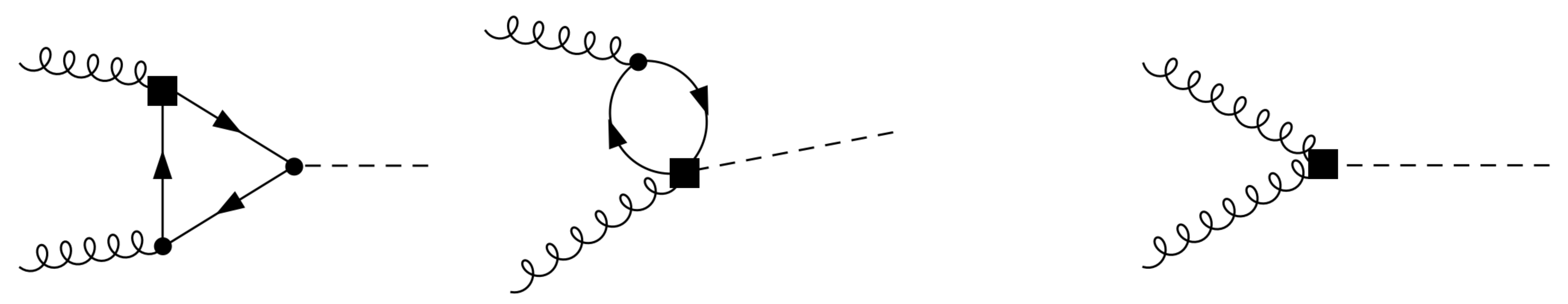}
\end{center}
\vspace{-2mm} 
\caption{\label{fig:diagrams2} Left:~Examples of one-loop diagrams for $gg \to h$ with a single insertion of $Q_{tG}$. Right:~The tree-level contribution to $gg \to h$ featuring an insertion of $Q_{HG}$. The plot styles follow the same conventions as in Figure~\ref{fig:diagrams1}.}
\end{figure}

Since the operator $Q_{G}$ mixes at one-loop and two-loop into $Q_{tG}$ and $Q_{HG}$, respectively, additional contributions to the $gg \to h$ process arise from the one-loop and tree-level matrix elements with insertions of these operators. Figure~\ref{fig:diagrams2} depicts representative graphs. The~bare one-loop amplitude with a $Q_{tG}$ insertion contains a local $1/\epsilon$ pole of UV origin, which is removed via $\overline{\rm MS}$ operator renormalization. The corresponding one-loop counterterm can be determined using the SMEFT anomalous dimensions provided in~\cite{Jenkins:2013wua,Alonso:2013hga}. Finally, the tree-level amplitude with a single insertion of $Q_{HG}$ yields a finite contribution to $gg \to h$. Further details on the matching calculation can be found in~Appendix~\ref{app:matching}. 

\section{Beta functions} 
\label{sec:betas}

The RG evolution of the Wilson coefficients $C_i$ is dictated by the beta functions:
\beq \label{eq:betai}
\frac{d C_i}{d \ln \mu} = \beta_i = \sum_{l=1}^\infty \frac{\beta_i^{(l)}}{(16 \pi^2)^l} \,. 
\eeq
The one-loop beta functions $\beta_i^{(1)}$ for the operators in~(\ref{eq:operators}) relevant to our study are
\beq \label{eq:beta1}
\begin{split}
\beta_{G}^{(1)} & = 15 \hspace{0.125mm} g_s^2 \hspace{0.25mm} C_{G} \,, \\[2mm]
\beta_{tG}^{(1)} & = \left ( -\frac{17}{3} \hspace{0.25mm} g_s^2 + \frac{15}{2} \hspace{0.25mm} y_t^2 \right ) C_{tG} + 9 \hspace{0.125mm} g_s^2 \hspace{0.25mm} y_t \hspace{0.25mm} C_{G} - 4 \hspace{0.125mm} g_s \hspace{0.125mm} y_t \hspace{0.25mm} C_{HG} \,, \\[2mm]
\beta_{HG}^{(1)} & = \Big ( -14 \hspace{0.125mm} g_s^2 + 6 \hspace{0.125mm} y_t^2 \Big ) \hspace{0.25mm} C_{HG} - 4 \hspace{0.125mm} g_s \hspace{0.125mm} y_t \hspace{0.25mm} C_{tG} \,.
\end{split}
\eeq
They have been computed in~\cite{Jenkins:2013wua,Alonso:2013hga}. Note that the coefficients proportional to $g_s^2$ in the beta functions $\beta_{G}^{(1)}$ and $\beta_{HG}^{(1)}$ depend on the number of active quark flavors. The given values are based on the assumption of six active quark flavors. Above, the dependence of $g_s$ and $y_t$ as well as the Wilson coefficients $C_i$ on the renormalization scale $\mu$ has been omitted.

As discussed in the previous section, the local $1/\epsilon$ pole that remains in our two-loop computation of the $gg \to h$ amplitude can be used to extract the unknown two-loop beta function in the SMEFT\footnote{Other two-loop anomalous dimensions in the SMEFT have been calculated in~\cite{Gorbahn:2016uoy,Bern:2020ikv,Jin:2020pwh,Jenkins:2023bls,DiNoi:2023ygk,Heinrich:2023rsd,DiNoi:2024ajj,Born:2024mgz,Duhr:2025zqw}.} that describes the mixing of $Q_{G}$ into~$Q_{HG}$. We~find 
\beq \label{eq:beta2}
\beta_{HG}^{(2)} = 207 \hspace{0.25mm} g_s^3 \hspace{0.25mm} y_t^2 \hspace{0.25mm} C_{G} \,.
\eeq
Upon transformation from the Warsaw to the Mainz basis, our result agrees with the two-loop SMEFT beta functions of~\cite{Born:2026xkr}, which were published after this work had been completed and thus provide an independent cross-check.

\section{Form factor} 
\label{sec:formfactor}

We now proceed to present the matching corrections to the $gg \to h$ process involving the Wilson coefficients of the operators~(\ref{eq:operators}). To establish our notation and normalization, we begin by giving the one-loop contribution to the $gg \to h$ form factor in the SM, induced by the top quark:
\beq \label{eq:gghoneloop}
G^{(1)} = -\frac{6 \hspace{0.125mm} x}{\left ( x - 1 \right )^2}+ \frac{3 \hspace{0.125mm} x \left ( x + 1 \right )^2}{\left ( x - 1 \right )^4} \hspace{0.25mm} H(0,0; x) \,,
\eeq
with 
\beq \label{eq:xtauFtri}
x = \frac{\sqrt{1 - \tau} - 1}{\sqrt{1 - \tau} + 1} \,, \qquad \tau = \frac{4 m_t^2}{m_h^2} \,.
\eeq
Here, $H(0,0; x)$ is an HPL of weight 2 that represents the~$\epsilon^0$~contribution to the massive one-loop triangle integral. The expressions for all HPLs appearing in our calculations in terms of logarithms, dilogarithms, and other well-known functions can be found in~Appendix~\ref{app:HPLs}. Note~that the HPL above depends on a single variable, $x$, known as the Landau variable, which is especially well-suited for describing two-scale Feynman integrals, such as those arising in our work. In the infinite top-quark mass limit,~i.e.,~as~$\tau \to \infty$, one has~$x \to 1$. In~this limit,~$G^{(1)} \to 1$, which fixes the normalization of the one-loop SM $gg \to h$ form~factor. 

The SMEFT corrections of interest to the $gg \to h$ form factor can be expressed as the sum of three separate contributions
\beq \label{eq:gghtwoloop}
G^{(2)} = \left ( \frac{g_s \hspace{0.125mm} y_t^2}{16 \pi^2} \hspace{0.125mm} F_{G} \hspace{0.25mm} C_{G} + \frac{y_t}{g_s} \hspace{0.125mm} F_{tG} \hspace{0.25mm} C_{tG} + \frac{16 \pi^2}{g_s^2} \hspace{0.125mm} F_{HG} \hspace{0.25mm} C_{HG} \right ) v^2 \,,
\eeq
where $v = 246.22 \, {\rm GeV}$ denotes the Higgs vacuum expectation value. 

The function $F_{G}$ encodes the two-loop corrections originating from $Q_{G}$. We obtain 
\bea \label{eq:FG}
\begin{split}
& F_{G}(x) = -\frac{6327}{8} + \frac{27 x}{\left ( x - 1 \right )^2}
 + \frac{27 \left ( x + 1 \right ) \left(17 x^2-36 x+17\right)}{2 \left ( x - 1 \right )^3} \, H(0; x) \\[2mm]
& \phantom{xx} + \frac{27 x \left(13 x^3-9 x^2-17 x+17\right)}{2 \left ( x - 1 \right )^4} \, H(0, 0; x)
 - \Bigg [ \, \frac{81}{2} + \frac{81}{x-1} \Bigg ] \, H(1, 0; x) \\[2mm]
& \phantom{xx} - \Bigg [ \, 216 + \frac{432}{x-1} \Bigg ] \, H(-1, 0; x)
 - \frac{27 x \left(4 x^2+3 x-3\right)}{2 \left ( x - 1 \right )^3} \, H(0,0,0;x) \\[2mm]
& \phantom{xx} - \Bigg [ \, 162 + \frac{1026 x}{\left ( x - 1 \right ) ^2} \Bigg ] \, H(1,0,0;x)
 + \Bigg [ \, 54 + \frac{135 x}{\left ( x - 1 \right ) ^2}\Bigg] \, H(0,1,0;x) \\[2mm]
& \phantom{xx} + \Bigg [ \, 54 + \frac{216 x}{\left ( x - 1 \right ) ^2} \Bigg ] \, H(0,0,1;x)
 - \frac{432 x}{\left ( x - 1 \right ) ^2} \, H(0,-1,0;x) \\[2mm]
& \phantom{xx} + \frac{27 x}{\left ( x - 1 \right ) ^2} \, H (x) - 54 B(x)
 - \Bigg [ \, \frac{297 \left ( x + 1 \right )}{2 \left ( x - 1 \right )}
 + \frac{297 x}{\left ( x - 1 \right ) ^2} \, H(0;x) \\[1mm]
& \phantom{xxxx} - \frac{27 x }{\left ( x - 1 \right ) ^2} \, H(0,0;x)
 - \frac{54 x}{\left ( x - 1 \right ) ^2} \, H(1,0;x) \Bigg ] \, H(0,1;1) \\[2mm]
& \phantom{xx} + \Bigg [ \, 216 + \frac{432 x}{\left ( x - 1 \right ) ^2} \Bigg ] \, H(0,0,1;1)
 + \frac{135 x}{2 \left ( x - 1 \right ) ^2} \, H(0,0,0,1;1) \\[2mm]
& \phantom{xx} - \Bigg [ \, 324 - \frac{27 \left ( x + 1 \right ) }{2 \left ( x - 1 \right ) } \, H(0;x)
 - \frac{27 x }{\left ( x - 1 \right ) ^2} \, H(0,0;x) \Bigg ] \, L_t - \frac{27}{2} \hspace{0.25mm} L_t^2 \,.
\end{split}
\eea 
In~(\ref{eq:FG}), $H(x)$ represents the following specific combination of HPLs of weight 4:
\beq \label{eq:H4}
\begin{split} 
H (x) & = 3 H(1,0,0,0;x) + 2 H(0,1,0,0;x) + H(0,0,1,0;x) \\[2mm] 
& \phantom{xx} + 4 H(1,1,0,0;x) + 2 H(1,0,1,0;x) \,. 
\end{split}
\eeq
Moreover, we have introduced the abbreviations
\beq \label{eq:BL}
B(x) = 1 - H \left (0; \frac{(1-x)^2}{x} - i \hspace{0.125mm} \delta \right ) \,, \qquad L_t = \ln \left ( \frac{\mu^2}{m_t^2} \right ) \,,
\eeq
where $B(x)$ results from the $\epsilon^0$ term in the massless one-loop on-shell bubble integral, with~$\delta$~being positive and infinitesimally small. Notice that the $i \hspace{0.125mm} \delta$ prescription is needed here for the correct analytic continuation of $H(0;x)$, which is merely a simple logarithm --- see~Appendix~\ref{app:HPLs}.

The function $F_{tG}$ in~(\ref{eq:gghtwoloop}) encapsulates the one-loop correction due to the insertion of the operator $Q_{tG}$. This contribution is explicitly given by:
\beq \label{eq:FtG}
F_{tG} = 3 \left [ 1 - \frac{x+1}{x - 1} \, H(0;x) -\frac{2 x }{\left ( x - 1 \right )^2} \, H(0,0;x) + 2 \hspace{0.125mm} L_t \right ] \,.
\eeq
This result agrees with the findings of~\cite{Grazzini:2016paz,Deutschmann:2017qum,Grazzini:2018eyk,DiNoi:2023ygk}. 

The tree-level contribution from $Q_{HG}$ is finally given by 
\beq \label{eq:FHG}
F_{HG} = 3 \,. 
\eeq

Before analyzing the RG flow of the contributions to $gg \to h$ production arising from the triple-gluon operator, we emphasize that in~(\ref{eq:gghtwoloop}), the Wilson coefficients $C_{G}$, $C_{tG}$ and~$C_{HG}$, as well as the strong coupling constant $g_s$ and the top-quark Yukawa coupling~$y_t$, are all understood to be evaluated at the renormalization scale $\mu$. 

\section{RG analysis} 
\label{sec:RG}

We start our study of the RGEs~(\ref{eq:betai}) by examining the following initial conditions
\beq \label{eq:CiLambda}
C_{G}(\Lambda) \neq 0 \,, \qquad C_{tG} (\Lambda) = 0 \,, \qquad C_{HG} (\Lambda) = 0 \,, 
\eeq
which corresponds to the extreme case of the hierarchy~(\ref{eq:hierarchy}), where at the high-energy scale~$\Lambda$, only the triple-gluon operator is generated through matching to a UV-complete BSM model. Inserting the beta functions~(\ref{eq:beta1}) and~(\ref{eq:beta2}) into the analytical results presented in~Appendix~\ref{app:RG}, we obtain the following simple results
\beq \label{eq:Cratios}
\begin{split}
\frac{C_{G} (\mu)}{C_{G} (\Lambda)} & \simeq 1 + \frac{g_s^2}{16 \pi^2} \left ( -\frac{15}{2} \hspace{0.25mm} L_\Lambda \right ) \,, \\[2mm]
\frac{C_{tG} (\mu)}{C_{G} (\Lambda)} & \simeq \frac{g_s^2 y_t}{16 \pi^2} \left ( -\frac{9}{2} \hspace{0.25mm} L_\Lambda \right ) \,, \\[2mm]
\frac{C_{HG} (\mu)}{C_{G} (\Lambda)} & \simeq \frac{g_s^3 y_t^2}{256 \pi^4} \left ( -\frac{9}{2} \hspace{0.25mm} L_\Lambda^2 - \frac{207}{2} \hspace{0.25mm} L_\Lambda \right ) \,, 
\end{split}
\eeq
where 
\beq \label{eq:LLambda}
L_\Lambda = \ln \left ( \frac{\Lambda^2}{\mu^2} \right ) \,.
\eeq
A few remarks seem warranted. The single logarithm in the evolution of $C_{G}$ results from the one-loop self-mixing of $Q_{G}$. For the running of $C_{tG}$, the single logarithm arises from the one-loop mixing of $Q_{G}$ into $Q_{tG}$. In the case of the evolution of $C_{HG}$, the double logarithm stems from a two-step mixing process,\footnote{The importance of logarithmic corrections resulting from a chain of operator mixings has been emphasized in several previous studies, including~\cite{Hisano:2012cc,Brod:2013cka,Cirigliano:2016njn,Cirigliano:2016nyn,Buras:2018gto,Panico:2018hal,Brod:2018pli,Bauer:2020jbp,Ardu:2021koz,Allwicher:2023aql,Brod:2023wsh,Biekotter:2023mpd, Garosi:2023yxg,Allwicher:2023shc,Stefanek:2024kds}.} namely $Q_{G}$ mixes into~$Q_{tG}$ at one loop, which subsequently mixes into $Q_{HG}$ at the same perturbative order. The single logarithm instead results from the direct two-loop mixing of $Q_{G}$ into~$Q_{HG}$. Notice that, from the viewpoint of RG improved perturbation theory, the double logarithm in the last line of~(\ref{eq:Cratios}) corresponds to a LL~effect, while the single logarithm represents a next-to-leading logarithmic~(NLL)~correction.

\begin{figure}[t!]
\begin{center}
\includegraphics[width=0.6\textwidth]{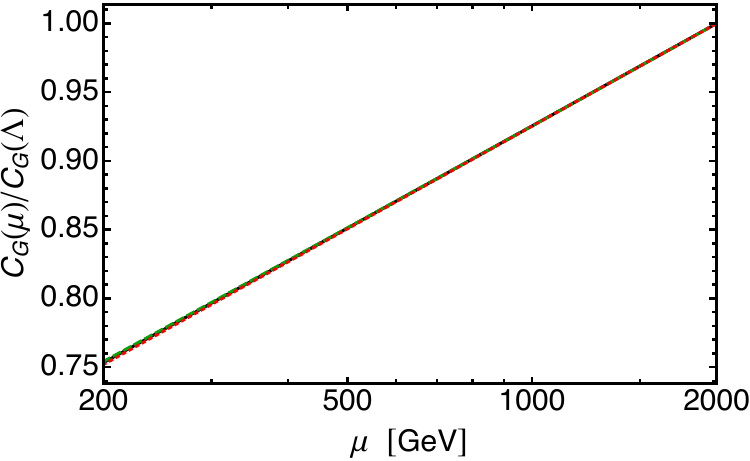}

\vspace{5mm} 

\includegraphics[width=0.6\textwidth]{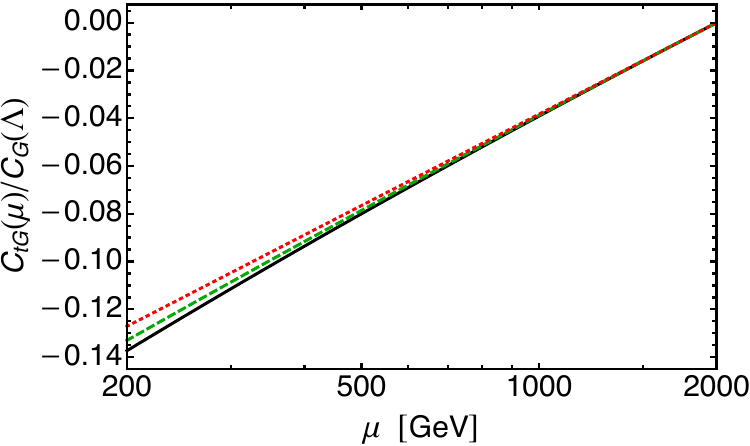}

\vspace{5mm} 

\includegraphics[width=0.6\textwidth]{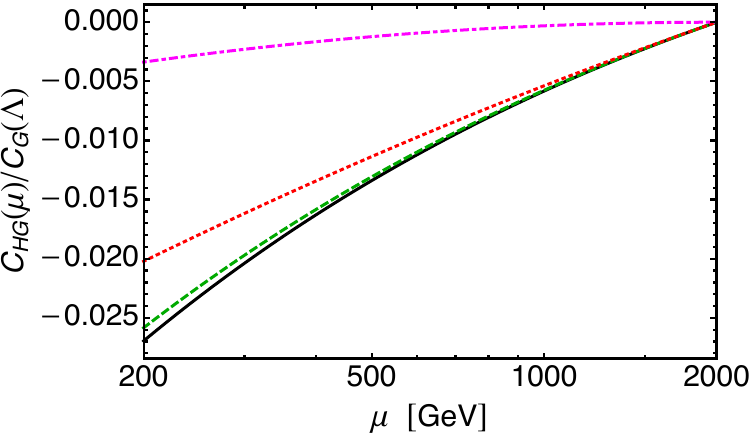}
\end{center}
\vspace{-2mm} 
\caption{\label{fig:running} Ratios $C_{G}(\mu)/C_{G}(\Lambda)$~(top), $C_{tG}(\mu)/C_{G}(\Lambda)$~(middle), and $C_{HG}(\mu)/C_{G}(\Lambda)$~(bottom) plotted as functions of the renormalization scale $\mu$. At the high-energy scale $\Lambda = 2 \, {\rm TeV}$, only the Wilson coefficient $C_{G} (\Lambda)$ is assumed to be nonzero. In the plots, the solid black curve depicts the numerical solution computed using {\tt DsixTools~2.0}, while the green dashed curve and red dotted curve represent the resummed analytic results and the logarithmically accurate results, respectively. The magenta dash-dotted curve in the bottom panel finally represents the LL-accurate result of the ratio $C_{HG}(\mu)/C_{G}(\Lambda)$. Consult the main text for additional explanations.}
\end{figure}

Before discussing the resummation of large logarithms, we emphasize the following important point. By inserting the logarithmic expressions for the ratios~(\ref{eq:Cratios}) of Wilson coefficients into~(\ref{eq:gghtwoloop}), it is a matter of simple algebra to demonstrate that the two-loop SMEFT corrections to the $gg \to h$ form factor become renormalization scale independent at ${\cal O} (g_s y_t^2)$. This occurs because the $L_t$-dependent terms in~(\ref{eq:FG}) to~(\ref{eq:FHG}) combine with the~$L_\Lambda$ contributions in~(\ref{eq:Cratios}), resulting in logarithms of the form $L = L_t + L_\Lambda = \ln \left (\Lambda^2/m_t^2 \right)$. The cancellation of renormalization scale dependence provides a strong consistency check for our computation. Indeed, the observation that different types of contributions --- incorporating matching and running effects at various perturbative orders --- must be included to ensure that observables remain free from unphysical scale dependencies is a general feature of EFTs at the quantum level. The case examined in this article, involving the three operators~(\ref{eq:operators}) that contribute to $gg \to h$ production, serves as a nontrivial yet insightful example within the SMEFT framework, highlighting this general aspect of RG improved perturbation theory. 

Using the RG formulas presented in~Appendix~\ref{app:RG} we can also resum the logarithms appearing in~(\ref{eq:Cratios}). We obtain 
\bea \label{eq:ratioC}
\begin{split}
\frac{C_{G} (\mu)}{C_{G} (\Lambda)} & \simeq \left ( \frac{\alpha_s (\mu)}{\alpha_s (\Lambda)} \right )^{-\frac{15}{14}} \,, \\[4mm]
\frac{C_{tG} (\mu)}{C_{G} (\Lambda)} & \simeq -\frac{9 \hspace{0.125mm} g_s^2 (\Lambda) \hspace{0.25mm} y_t (\Lambda)}{16 \pi^2} \left [ \frac{14011}{935} \hspace{0.25mm} \big ( \alpha_s (\mu) \big )^{\frac{3}{50}} -\frac{10789}{4554} \hspace{0.25mm} \big( \alpha_s (\mu) \big )^{-\frac{938}{1327}} \right ] \,, \\[4mm]
\frac{C_{HG} (\mu)}{C_{G} (\Lambda)} & \simeq -\frac{9 \hspace{0.125mm} g_s^3 (\Lambda) \hspace{0.25mm} y_t^2 (\Lambda)}{128 \pi^4} \left [ \frac{69439}{77} \hspace{0.25mm} \big ( \alpha_s (\mu) \big )^{\frac{21}{29}} + \frac{1803}{83} \hspace{0.25mm} \big( \alpha_s (\mu) \big )^{-\frac{187}{222}} - \frac{9421}{35} \hspace{0.25mm} \big( \alpha_s (\mu) \big )^{-\frac{4}{53}} \right ] \\[2mm]
& \phantom{xx} -\frac{207 \hspace{0.125mm} g_s^3 (\Lambda) \hspace{0.25mm} y_t^2 (\Lambda)}{256 \pi^4} \left [ \frac{98329}{981} \hspace{0.25mm} \big ( \alpha_s (\mu) \big )^{\frac{470}{649}} -\frac{12317}{480} \hspace{0.25mm} \big( \alpha_s (\mu) \big )^{\frac{138}{875}} \right ] \,. 
\end{split}
\eea
Here, $\alpha_s = g_s^2/(4 \pi)$, and the results for the second and third ratios of Wilson coefficients use $g_s (\Lambda) \simeq 1.065$ and $y_t (\Lambda) \simeq 0.854$, valid for $\Lambda = 2 \, {\rm TeV}$. The numerical values quoted were obtained using {\tt DsixTools~2.0}~\cite{Fuentes-Martin:2020zaz}. Note that the double-logarithmic contribution, representing an LL term in the running of $C_{HG}$, arises from the first line of the third expression in~(\ref{eq:ratioC}), whereas the single-logarithmic correction, corresponding to an NLL term, originates from the second line.

In Figure~\ref{fig:running}, we present a comparison between the numerical solutions for the ratios $C_{G}(\mu)/C_{G}(\Lambda)$, $C_{tG}(\mu)/C_{G}(\Lambda)$, and $C_{HG}(\mu)/C_{G}(\Lambda)$ of Wilson coefficients and their corresponding approximations in~(\ref{eq:Cratios}) and (\ref{eq:ratioC}). The numerical solutions, shown as solid black curves, are obtained using {\tt DsixTools~2.0} through direct numerical integration of the RGEs. The version of {\tt DsixTools~2.0} used has been modified to include the two-loop beta function~$\beta_{HG}^{(2)}$ from~(\ref{eq:beta2}). The green dashed curves depict the resummed analytic results, while the red dotted curves represent the logarithmically accurate results. In the case of $C_{HG}(\mu)/C_{G}(\Lambda)$, the LL-accurate result is finally shown as a magenta dash-dotted curve. We further note that the resummed analytic results presented were obtained using the LO QCD approximation for the running strong coupling constant, as given in~(\ref{eq:alphasLORGE}). This~choice is justified by the fact that the LO QCD running of $\alpha_s (\mu)$ was used in deriving the approximations in~(\ref{eq:ratioC}). 

The first notable observation from Figure~\ref{fig:running} is that the expressions in~(\ref{eq:ratioC}) provide accurate approximations to the exact numerical results across all three cases studied. The~small discrepancies observed can be attributed to the inclusion of higher-order corrections in the evolution of $g_s$ and $y_t$ within {\tt DsixTools~2.0}, which are not accounted for in our approximations. We note that using the \texttt{DsixTools~2.0} prediction for $\alpha_s (\mu)$ would result in an almost perfect agreement between the resummed analytic results and the exact numerical results. In~the case of $C_{G}(\mu)/C_{G}(\Lambda)$ and $C_{tG}(\mu)/C_{G}(\Lambda)$, we find that the resummation of logarithmic effects through the RGEs has only a minor numerical impact. However, for~$C_{HG}(\mu)/C_{G}(\Lambda)$, resummation plays a more significant role. Consequently, the ratio given in~(\ref{eq:Cratios}), which includes both double-logarithmic and single-logarithmic terms, is a less accurate approximation. This is due to the large numerical coefficient in the two-loop beta function~(\ref{eq:beta2}). Therefore, the LL-accurate result for $C_{HG}(\mu)/C_{G}(\Lambda)$ is also not a good approximation. The findings above show that the dominant RG effects arise from operator mixings proportional to powers of $g_s$ or $y_t$, as well as the evolution of these parameters under the RG flow. As discussed in this section and in~Appendix~\ref{app:RG}, corrections of this nature can be resummed, resulting in relatively compact analytic expressions.
 
\section{Phenomenology} 
\label{sec:pheno}

Utilizing the form factors given in~(\ref{eq:gghoneloop}) and~(\ref{eq:gghtwoloop}), the Higgs production cross section in the $gg \to h$ process, incorporating the effects of the SMEFT operators introduced in~(\ref{eq:operators}), can be expressed as
\beq \label{eq:higgsXs}
\sigma \left (gg \to h \right ) = \frac{\alpha_s^2}{576 \hspace{0.125mm} \pi} \hspace{0.25mm} \frac{1}{v^2} \hspace{0.5mm} \big | G^{(1)} + G^{(2)} \big |^2 \,.
\eeq
To linear order in the Wilson coefficients, the resulting modification of the signal strength~$\mu_{gg}$ relative to the SM in the $gg \to h$ channel is therefore given by
\beq \label{eq:kappag}
\delta \mu_{gg} = 2 \hspace{0.125mm} \delta \kappa_{g} = \frac{2 \hspace{0.25mm} {\rm Re} \left ( G^{(1)} G^{(2) \hspace{0.25mm} \ast} \right )}{ \big | G^{(1)} \big |^2} \,. 
\eeq

As a concrete example, let us derive the current constraint on the high-scale Wilson coefficient of $Q_{G}$ using the existing LHC measurement of the effective Higgs-gluon coupling modifier~$\kappa_g$. Taking $m_t = 162.5 \, {\rm GeV}$ and $m_h = 125.2 \, {\rm GeV}$ from the Particle~Data~Group~\cite{ParticleDataGroup:2024cfk} as input parameters, and setting $\Lambda = 2 \, {\rm TeV}$, we obtain the following numerical result 
\beq \label{eq:deltakappag}
\frac{\delta \kappa_{g}}{v^2} = -6.5 \left [1 + 0.034 \ln \left ( \frac{\mu^2}{m_h^2} \right ) - 0.0073 \ln^2 \left ( \frac{\mu^2}{m_h^2} \right ) \right ] C_{G} (\Lambda) \,, 
\eeq
for the modification of~$\kappa_g$ in the BSM scenario~(\ref{eq:CiLambda}). The numerical values of the low-scale Wilson coefficients entering~(\ref{eq:deltakappag}) have been obtained using {\tt DsixTools~2.0}. Regarding the above result, we stress that the sensitivity of~$\delta \kappa_{g}$ to~$C_{G}(\Lambda)$ is reduced by a destructive interference between the contributions of the form factors~$F_{G}$ and~$F_{tG}$ to~(\ref{eq:gghtwoloop}) and that of~$F_{HG}$. This cancellation, however, is not coincidental but stems from the renormalization scale independence of $G^{(2)}$ up to~NLL~accuracy.

At the $68\%$ confidence level (CL), the ATLAS collaboration reports the following constraint on~$\kappa_g$:
\beq \label{eq:kappagATLASRunII}
\kappa_{g} = 0.949^{+0.073}_{-0.067} \,,
\eeq
which is based on the full LHC Run~II~dataset~\cite{ATLAS:2022vkf}. By applying~(\ref{eq:deltakappag}) with $\mu = m_h$, we translate~(\ref{eq:kappagATLASRunII}) into the $68\%$~CL bound
\beq \label{eq:ourlimit}
C_{G} (\Lambda) = \frac{[-0.055, 0.30]}{{\rm TeV}^2} \,.
\eeq
Limits on the Wilson coefficient~$C_{G}$ have also been derived in~\cite{Ghosh:2014wxa,Krauss:2016ely,Hirschi:2018etq,Goldouzian:2020wdq,Bardhan:2020vcl,Ellis:2020unq}. The strongest bound to date, $|C_{G}| < 0.031 \, {\rm TeV}^{-2}$ at $95\%$~CL, was obtained in~\cite{Goldouzian:2020wdq} from LHC dijet angular distributions. Although~(\ref{eq:ourlimit}) is weaker than this limit, the following points should be kept in mind. The constraints on the triple-gluon operator from multijet production are driven by terms proportional to~$|C_{G}|^2$, while the linear terms remain negligible, even for more than two jets~\cite{Krauss:2016ely,Hirschi:2018etq}. This~feature arises from the strong energy enhancement of the quadratic SMEFT corrections relative to the SM background. In contrast, the limit~(\ref{eq:ourlimit}) arises at linear order and involves only virtualities well below~$\Lambda$. From the standpoint of the robustness and validity of the EFT expansion, the bound on~$C_{G}$ derived here is thus more reliable than the limits obtained from multijet production. We therefore conclude that LHC measurements of the $gg \to h$ production cross section provide complementary probes of anomalous gluon~dynamics.

\section{Conclusions} 
\label{sec:conclusions}

This article is part of the SMEFT precision program~(see,~e.g.,~\cite{Haisch:2022nwz,DiNoi:2023ygk,Gauld:2023gtb,Heinrich:2023rsd,DiNoi:2024ajj,Born:2024mgz,Haisch:2024wnw,Bonetti:2025hnb,Gorbahn:2016uoy,Bern:2020ikv,Jin:2020pwh,Jenkins:2023bls,Deutschmann:2017qum}), which aims at improving the precision of SMEFT calculations beyond the one-loop level. Specifically, we performed a comprehensive two-loop analysis of the contributions to Higgs production via $gg \to h$ from the triple-gluon operator, which captures anomalous gluon~dynamics. Working in the broken phase of the theory, we determined the exact dependence of the relevant amplitudes on the Higgs and top-quark masses. We also provided a concise discussion of the renormalization procedure that guarantees the UV finiteness of the resulting two-loop $gg \to h$ form factor. Additional details regarding real radiation corrections and the matching procedure at one and two loops can be found in Appendix~\ref{app:realradiation} and Appendix~\ref{app:matching}, respectively. It turns out that the renormalization of the calculated two-loop amplitudes requires an unknown two-loop SMEFT anomalous dimension, which we computed as a byproduct.

By analyzing the RG flow of the relevant SMEFT Wilson coefficients, we have demonstrated that achieving a renormalization scale independent result for the $gg \to h$ production cross section at next-to-leading order in the SMEFT, in the case under study, requires accounting for three types of contributions: $(i)$~two-loop matching contributions involving the triple-gluon operator, $(ii)$~one-loop matching contributions associated with the chromomagnetic top-quark dipole operator, and~$(iii)$ tree-level matching contributions arising from insertions of the Higgs-gluon operator. We emphasized that the necessity of including distinct types of contributions, involving matching and running effects at different perturbative orders, to achieve renormalization scale independent results is a general feature of SMEFT calculations beyond the tree level. The case studied in this article provides a nontrivial yet instructive two-loop example that emphasizes this general aspect of RG improved perturbation~theory. 

Utilizing the SMEFT RGEs, we then resummed the large logarithms that appear in the $gg \to h$ prediction. Specifically, we derived simple analytic expressions and compared them with the exact results obtained from {\tt DsixTools~2.0}~\cite{Fuentes-Martin:2020zaz} via direct numerical integration of the RGEs. The presented expressions provide accurate approximations to the exact numerical results in all three cases analyzed. We found that resumming logarithmic effects has a limited numerical impact on our results. The only partial exception is the Wilson coefficient of the Higgs-gluon operator, which receives sizeable NLL corrections due to the newly calculated two-loop beta function. Our~findings demonstrate that, in typical phenomenological applications, the dominant SMEFT RG effects stem from operator mixings that scale with powers of the strong coupling constant or the top-quark Yukawa coupling, along with the RG evolution of these parameters. A general framework for resumming SMEFT corrections of this type is outlined in Appendix~\ref{app:RG}. Our approach is based on the pedagogical discussion provided in~\cite{Buras:2018gto}.

We also explored the phenomenological implications of our two-loop calculation. To~this end, we derived a numerical expression for the modification of the signal strength in the $gg \to h$ process. The derived formula was subsequently used to obtain constraints on the high-scale Wilson coefficient of the triple-gluon operator. We found that the resulting bound is weaker than the existing constraints from multijet production~\cite{Krauss:2016ely,Hirschi:2018etq,Goldouzian:2020wdq}. However,~when comparing these constraints, it is important to remember that the latter bounds stem from the strong enhancement of the quadratic contributions in the Wilson coefficient within the high-energy tails of multijet distributions, while the $gg \to h$ process probes the linear corrections at virtualities well below the UV cut-off. From the perspective of the robustness and validity of the EFT expansion, the limit derived here is therefore more reliable than those obtained from multijet production. Given this complementarity, we believe that constraints on the triple-gluon operator from $gg \to h$ should be included in global SMEFT~analyses. The compact analytical and numerical expressions presented in this study should prove useful in this context.

\acknowledgments{UH expresses gratitude to Marco~Niggetiedt, Luc~Schnell, and Ben~Stefanek for their collaborations on related subjects. He would also like to thank Gino~Isidori and Zach~Polonsky for inviting him to ``ZPW2025: Particle Physics from Low to High Energies''. A part of this workshop was dedicated to celebrating Daniel Wyler’s 75th birthday, recognizing him as one of the two pioneers of the SMEFT~\cite{Buchmuller:1985jz}, among other accomplishments. Happy~birthday,~Daniel! This work has benefited greatly from the workshop’s relaxed atmosphere and the long train ride between Munich and Zurich, and back again. We are also grateful to the anonymous referee for their insightful question regarding IR divergences, which led to improvements to our work, now included in Appendix~\ref{app:realradiation}. Special thanks go to Giulia Zanderighi for carefully reading the revised parts of this article and offering useful feedback. UH also thanks Giuseppe Ventura for pointing out a mistake in~(\ref{eq:FG}) in the published version of this article and for verifying the new result.} The Feynman diagrams in this article were created using {\tt FeynArts}~\cite{Hahn:2000kx}.

\begin{appendix}

\section{Real radiation}
\label{app:realradiation}

\begin{figure}[t!]
\begin{center}
\includegraphics[width=0.575\textwidth]{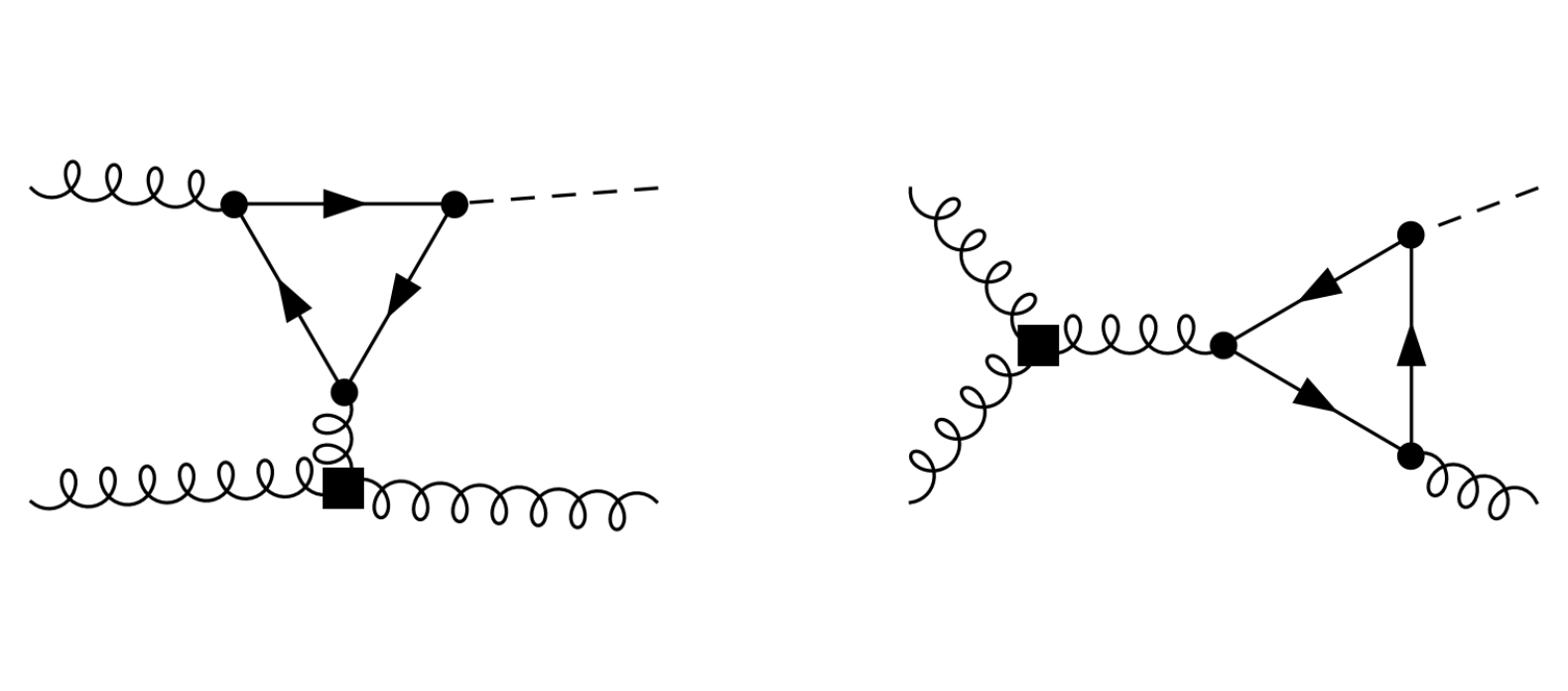}
\end{center}
\vspace{-8mm} 
\caption{\label{fig:diagrams3} Examples of one-loop contributions to the $gg \to hg$ process with a $Q_G$ operator insertion, as defined in~(\ref{eq:operators}). The plot styles follow the conventions established in~Figure~\ref{fig:diagrams1}.} 
\end{figure}

In this appendix, we discuss the effect of real radiation on the calculation of the form factor~$F_{G}$ given in~(\ref{eq:FG}), which accounts for the two-loop corrections from the operator~$Q_{G}$. Representative Feynman diagrams for the process $gg \to h g$ are shown in~Figure~\ref{fig:diagrams3}. Before delving into the computation of these diagrams, we recall that in the SM, to calculate the~${\cal O} (\alpha_s)$ corrections to the inclusive production of the Higgs via gluon-gluon fusion, both the real contributions from $g g \to g h$ and the virtual loop corrections to~$g g \to h$ are required, as only their sum is IR finite. A careful reader, such as the referee, might therefore be puzzled as to why IR divergences were not addressed at all in Section~\ref{sec:calculation} of the original version of this manuscript. We will now seek to rectify this omission.

We start by noting that factorization ensures that all IR singular contributions can be expressed in terms of the Born amplitudes, which, in our case, corresponds to the $gg \to h$ amplitude. This holds true regardless of whether the $gg \to h$ amplitude is calculated exactly or in the infinite top-quark mass limit. Without loss of generality, we can thus work in the limit where $m_t^2/m_h^2 \to \infty$, which simplifies the subsequent discussion. The interference term between the matrix elements of $g(p_1) g(p_2) \to h(p_3) g(p_4)$ with an insertion of $Q_{G}$ and the corresponding SM contribution results, in the infinite top-quark mass limit, in the following spin and color averaged contribution:
\beq \label{eq:real2}
{\cal R} = 2 \hspace{0.5mm} {\rm Re } \hspace{0.25mm} \big \langle {\cal M}_{Q_{G}} \big | {\cal M}_{\rm SM} \big \rangle = \frac{36 \hspace{0.125mm} g_s}{\pi} \left ( 3 \hspace{0.125mm} m_h^4 + s^2 + t^2 + u^2 \right ) C_{G} \hspace{0.375mm} \sigma_{\cal B} \,.
\eeq
Here, $s = (p_1 + p_2)^2$, $t = (p_1 - p_3)^2$, and $u = (p_1 - p_4)^2$ are the usual Mandelstam variables, satisfying the relation $s + t + u = m_h^2$, and 
\beq \label{eq:sigmaB}
\quad \sigma_{\cal B} = \frac{\alpha_s^2}{576 \hspace{0.125mm} \pi} \frac{1}{v^2} \,,
\eeq
represents the inclusive Born-level cross section for $gg \to h$ in the SM, evaluated in the infinite top-quark mass limit.

The cross section associated with the real contribution from $gg \to h g$ is obtained by integrating~(\ref{eq:real2}) over the relevant phase space:
\beq \label{eq:sigmaR}
\sigma_{\cal R} = \frac{1}{2 s} \int \! d\Phi_2 \, {\cal R} \,. 
\eeq
A useful parameterization of the two-particle phase space appearing in~(\ref{eq:sigmaR}) in $d = 4 - 2 \epsilon$ dimensions is given by
\beq \label{eq:dPhi2}
d\Phi_2 = \frac{1}{8 \pi} \left ( \frac{4 \pi}{s} \right )^\epsilon \frac{1}{\Gamma \left ( 1 - \epsilon \right )} \left ( 1 - \frac{m_h^2}{s} \right )^{1 - 2 \epsilon} \int_{0}^{1} \! d\alpha \, \alpha^{-\epsilon} \left ( 1 - \alpha \right )^{-\epsilon} \,, 
\eeq
and the Mandelstam variables $t$ and $u$ can be expressed in terms of $\alpha$ as follows: 
\beq \label{eq:tuparametrization}
t = - s \left ( 1 - \frac{m_h^2}{s} \right ) \left ( 1 - \alpha \right ) \,, \qquad u = - s \left ( 1 - \frac{m_h^2}{s} \right ) \alpha \,. 
\eeq

Utilizing~(\ref{eq:sigmaR}), (\ref{eq:dPhi2}), and (\ref{eq:tuparametrization}), it is straightforward to integrate~(\ref{eq:real2}). We obtain the IR-finite result
\beq \label{eq:sigmaRfinal}
\sigma_{\cal R} = \frac{3 g_s}{4 \pi^2} \hspace{0.75mm} s \left (1 - \frac{m_h^2}{s} \right ) \left ( \frac{11 \hspace{0.125mm} m_h^4}{s^2} - \frac{4 \hspace{0.125mm} m_h^2}{s} + 5 \right ) \hspace{0.5mm} C_{G} \hspace{0.375mm} \sigma_{\cal B} \,, 
\eeq
with the expression for the Born-level cross section $\sigma_{\cal B}$ given in~(\ref{eq:sigmaB}). Notice that in the limit $s \to m_h^2$, the real-emission contribution $\sigma_{\cal R}$ vanishes. Therefore, the form factor~$F_G$ in~(\ref{eq:FG}) receives contributions exclusively from virtual corrections, while real emissions linked to the triple-gluon operator $Q_G$ do not contribute. It follows that the sum of the two-loop Feynman diagrams in~Figure~\ref{fig:diagrams1} represents the full bare IR-finite contribution to the $gg \to h$ amplitude, and that the $1/\epsilon^2$ and $1/\epsilon$ poles that it contains are purely UV in nature, as stated in~Section~\ref{sec:calculation}. After subtracting the corresponding counterterms, the remaining $1/\epsilon$~pole~can thus be identified with the two-loop beta function given in~(\ref{eq:beta2}). Although somewhat unexpected, we point out that the IR finiteness of SMEFT real radiation contributions --- which are IR divergent in the SM --- has been previously observed, for example in the case of the $h \to b \bar b$ decay~\cite{Haisch:2022nwz}. 

\section{Matching procedure}
\label{app:matching}

In this appendix, we provide a more detailed explanation of the matching procedure that yields the results~(\ref{eq:FG}) and (\ref{eq:FtG}). We start by discussing the one-loop matching that leads to the term~$F_{tG}$ in the two-loop form factor $G^{(2)}$ as presented in~(\ref{eq:gghtwoloop}). In this case, the relevant matching equation reads:
\beq \label{eq:onematch} 
K^{(1)} = A^{(1)} - \frac{Z_{tG,HG}^{(1),1}}{\epsilon} \,.
\eeq
Here, $A^{(1)}$ denotes the bare one-loop amplitude resulting from single insertions of $Q_{tG}$, normalized to the tree-level $gg \to h$ matrix element of $Q_{HG}$, with an overall loop factor of~$(16 \pi^2)^{-1}$ factored out. This amplitude arises from the Feynman diagrams shown on the left-hand side of Figure~\ref{fig:diagrams2}. The symbol~$Z_{tG,HG}^{(1),1}$ represents the one-loop $1/\epsilon$ pole of the~$Z$~factor, which accounts for the mixing of $Q_{tG}$ into~$Q_{HG}$. Specifically, it is given by
\beq \label{eq:firstZ}
Z_{tG,HG}^{(1),1} = 2 \hspace{0.125mm} g_s \hspace{0.125mm} y_t \,.
\eeq
Observe that this $Z$~factor determines the coefficient of $C_{tG}$ in the one-loop beta function~$\beta^{(1)}_{HG}$ given in~(\ref{eq:beta1}). By evaluating the bare one-loop amplitude $A^{(1)}$ up to order $\epsilon$ in the Laurent series and carrying out the matching, we obtain:
\beq \label{eq:Kdecomp}
K^{(1)} = g_s \hspace{0.125mm} y_t \left ( K^{(1)}_0 + \epsilon \hspace{0.25mm} K^{(1)}_\epsilon \right ) \,, 
\eeq
where 
\beq \label{eq:KK}
\begin{split}
K^{(1)}_0 & = 1 - \frac{x + 1}{x - 1} \, H(0;x) -\frac{2 x }{\left ( x - 1 \right )^2} \, H(0,0;x) + 2 \hspace{0.125mm} L_t \,, \\[4mm]
K^{(1)}_\epsilon & = 1 - \frac{x + 1}{x - 1} \, H(0;x) - \frac{x^2 - 2 x - 1}{\left ( x - 1 \right )^2} \, H(0,0;x) + \frac{2 \left (x+1 \right )}{x-1} \, H(-1,0;x) \\[2mm]
& \phantom{xx} -\frac{2 x}{\left ( x - 1 \right )^2} \, H(0,0,0;x) + \frac{4 x}{\left ( x - 1 \right )^2} \, H(0,-1,0;x) \\[2mm]
& \phantom{xx} + \left [ \frac{2 x}{x-1} + \frac{2 x}{\left ( x - 1 \right )^2} \, H(0;x) \right ] H(0,1;1) + \frac{6 x}{\left ( x - 1 \right )^2} \, H(0,0,1;1) \\[2mm]
& \phantom{xx} + \left [ 1 - \frac{x+1}{x - 1} \, H(0;x) -\frac{2 x }{\left ( x - 1 \right )^2} \, H(0,0;x) \right ] L_t + L_t^2 \,.
\end{split}
\eeq
Notice that $K^{(1)}_0$ matches the result in (\ref{eq:FtG}) apart from an overall factor of 3, which cancels out when considering that the tree-level $gg \to h$ matrix element of $Q_{HG}$, in the normalization of~(\ref{eq:gghtwoloop}), is also 3. See~(\ref{eq:FHG}).

The matching equation required to determine the term~$F_{G}$ in the two-loop form factor~$G^{(2)}$ from~(\ref{eq:gghtwoloop}) is given by:
\beq \label{eq:twomatch} 
K^{(2)} = A^{(2)}- \Bigg ( \frac{Z_{G,HG}^{(2),2}}{\epsilon^2} + \frac{Z_{G,HG}^{(2),1}}{\epsilon} \Bigg ) - \frac{Z_{G,tG}^{(1),1}}{\epsilon} \hspace{0.5mm} K^{(1)} \,.
\eeq
The symbol $A^{(2)}$ represents the bare two-loop amplitude with a $Q_{G}$ insertion, normalized to the tree-level $gg \to h$ matrix element of $Q_{HG}$, with~$(16 \pi^2)^{-2}$ factored out. It arises from the graphs displayed in~Figure~\ref{fig:diagrams1}. The two-loop $Z$~factors in~(\ref{eq:twomatch}) are given by
\beq \label{eq:twoZs}
Z_{G,HG}^{(2),2} = \frac{Z_{G,tG}^{(1),1} \hspace{0.5mm} Z_{tG,HG}^{(1),1}}{2} \,, \qquad
Z_{G,HG}^{(2),1} = -\frac{207}{4} \hspace{0.5mm} g_s^3 \hspace{0.125mm} y_t^2 \,.
\eeq
Notably, the two-loop $1/\epsilon^2$ pole can be expressed as a product of one-loop $1/\epsilon$ poles, reflecting the locality of UV divergences (see,~e.g.,~\cite{Chetyrkin:1997fm,Gambino:2003zm}). The two-loop $1/\epsilon$ pole, on the other hand, is related to the new beta function~$\beta_{HG}^{(2)}$ in~(\ref{eq:beta2}). The value of the $Z$~factor~$Z_{tG,HG}^{(1),1}$ is given in~(\ref{eq:firstZ}), while the one-loop $1/\epsilon$ pole of the $Z$~factor describing the mixing of $Q_G$ into $Q_{tG}$ reads: 
\beq \label{eq:secondZ}
Z_{G, tG}^{(1),1} = -\frac{9}{2} \hspace{0.5mm} g_s^2 \hspace{0.125mm} y_t \,.
\eeq
This $Z$~factor is related to the coefficient of $C_{G}$ in the one-loop beta function~$\beta^{(1)}_{tG}$ from~(\ref{eq:beta1}). Notice that in~(\ref{eq:twomatch}), the order $\epsilon$ term of $K^{(1)}$, namely $K^{(1)}_\epsilon$, yields a finite contribution when multiplied with the $1/\epsilon$ factor proportional to $Z_{G, tG}^{(1),1}$. By combining the bare two-loop amplitude $A^{(2)}$ with the counterterms specified in~(\ref{eq:twomatch}), all UV poles cancel, yielding a finite expression for $K^{(2)}$. Up to an overall normalization factor, this result matches the expression for $F_{G}$ as given in~(\ref{eq:FG}).

\section{HPL formulas}
\label{app:HPLs}

The HPLs in this article can all be rewritten in terms of logarithms, dilogarithms, and similar functions using the {\tt HPL} package~\cite{Maitre:2005uu}. To make our paper self-contained, we provide all relevant formulas below.

The HPLs up to weight 3 that feature in our calculations are given by: 
\beq \label{eq:HPLs1} 
\begin{split}
H(0;x) & = \ln \left ( x \right ) \,, \\[2mm]
H(0,0;x) & = \frac{1}{2} \ln^2 \left ( x \right ) \,, \\[2mm]
H(1,0;x) & = -\ln \left (x \right ) \ln \left ( 1 - x \right ) - \text{Li}_2 \left (x \right ) \,, \\[2mm]
H(-1,0;x) & = \ln \left (x \right ) \ln \left ( 1 + x \right ) + \text{Li}_2 \left (-x \right ) \,, \\[2mm]
H(0,0,0;x) & = \frac{1}{6} \ln^3 \left (x \right ) \,, \\[2mm]
H(1,0,0;x) & = -\frac{1}{2} \ln \left ( 1 - x \right ) \ln^2 \left ( x \right ) - \ln \left ( x \right ) \text{Li}_2 \left (x \right ) + \text{Li}_3 \left (x \right ) \,, \\[2mm]
H(0,1,0;x) & = \ln \left ( x \right ) \text{Li}_2 \left ( x \right ) - 2 \hspace{0.125mm} \text{Li}_3 \left ( x \right ) \,, \\[2mm]
H(0,0,1;x) & = \text{Li}_3 \left ( x \right ) \,, \\[2mm]
H(0,-1,0;x) & = -\ln \left ( x \right ) \text{Li}_2 \left ( -x \right ) + 2 \hspace{0.125mm} \text{Li}_3 \left ( -x \right ) \,.
\end{split}
\eeq
Here, $\text{Li}_2 \left ( x \right )$ and $\text{Li}_3 \left ( x \right )$ represent the dilogarithm and trilogarithm, respectively.

HPLs of weight 4 appear in~(\ref{eq:FG}) only in the specific combination~(\ref{eq:H4}). Expressed through commonly known functions, we find 
\bea \label{eq:HPLs2} 
\begin{split}
H (x) & = \ln^2 \left ( x \right ) \ln^2 \left ( 1 - x \right ) - \frac{1}{2} \ln^3 \left ( x \right ) \ln \left ( 1 - x \right ) + 2 \ln \left ( x \right ) \ln \left ( 1 - x \right ) \text{Li}_2 \left (x \right ) \\[2mm]
& \phantom{x} - \frac{1}{2} \ln^2 \left ( x \right ) \text{Li}_2 \left (x \right ) + \big [ \text{Li}_2 \left (x \right ) \big ]^2 \,. 
\end{split}
\eea

The HPLs with argument 1 required in this article are
\beq \label{eq:HPLs3} 
H(0,1;1) = \frac{\pi^2}{6} \,, \qquad H(0,0,1;1) = \zeta(3) \,, \qquad H(0,0,0,1;1) = \frac{\pi^4}{90} \,,
\eeq
with $\zeta(3) \simeq 1.20206$, the value of the Riemann zeta function at argument 3.

\section{RG formulas}
\label{app:RG}

In this appendix, we present approximate analytical solutions for the coupled RGEs discussed in our article. Our approach follows, to some extent, the discussion in~\cite{Buras:2018gto}.

To derive approximate analytical solutions to~(\ref{eq:betai}), we begin by considering the RGE for the strong coupling constant $\alpha_s (\mu)$. At leading order (LO) in QCD, it is given by
\beq \label{eq:LOalphas} 
\frac{d \hspace{0.125mm} \alpha_s}{d \ln \mu} = -\beta_0 \hspace{0.25mm} \frac{\alpha_s^2}{2 \pi} \,, \qquad \beta_0 = 11 - \frac{2}{3} \hspace{0.125mm} N_F \,, 
\eeq
where $\beta_0$ denotes the LO coefficient of the QCD beta function, while $N_F$ represents the number of active quark flavors at the renormalization scale $\mu$. The solution to~(\ref{eq:LOalphas}) reads 
\beq \label{eq:alphasLORGE}
\alpha_s (\mu) \simeq \frac{\alpha_s (\mu_0)}{1 - \frac{\alpha_s (\mu_0)}{4 \pi} \hspace{0.25mm} \beta_0 \ln \left ( \frac{\mu_0^2}{\mu^2} \right )} \,.
\eeq

The beta functions in~(\ref{eq:beta1}) and~(\ref{eq:beta2}) depend not only on $\alpha_s$ but also on $y_t$. The~logarithms associated with the running of the top-quark Yukawa coupling can be resummed by using the following RGE~\cite{Cheng:1973nv,Machacek:1983fi}:
\beq \label{eq:RGEy1}
\frac{d \hspace{0.125mm} y_t}{d \ln \mu} = - \frac{y_t}{16 \pi^2} \left ( 4 \pi \gamma_m^0 \alpha_s - \frac{9}{2} \hspace{0.25mm} y_t^2 \right ) \,, \qquad \gamma_m^0 = 8 \,. 
\eeq
Here, $\gamma_m^0$ denotes the LO anomalous dimension of the quark mass in QCD. On the right-hand side of the differential equation, we have included the numerically dominant corrections proportional to $\alpha_s$ and $y_t^2$. By utilizing~(\ref{eq:LOalphas}) as well as the approximate scale independence of $y_t^2 (\mu)/\alpha_s (\mu) = {\rm const.}$~\cite{Pendleton:1980as,Hill:1980sq,Buras:2018gto}, we obtain the following solution
\beq \label{eq:ysol}
\frac{y_t (\mu)}{y_t (\mu_0)} \simeq \eta^{\frac{y}{b}} \,, 
\eeq
with 
\beq \label{eq:gendef}
\eta = \frac{\alpha_s (\mu)}{\alpha_s (\mu_0)} \,, \qquad b = 8 \pi \beta_0 \,, \qquad y = 4 \pi \gamma_m^0 \left ( 1 - \frac{9}{{8 \pi \gamma_m^0}} \hspace{0.25mm} \frac{y_t^2 (\mu_0)}{\alpha_s (\mu_0)} \right ) \,.
\eeq

The first type of operator mixing relevant to our analysis is the self-mixing of an operator $Q_i$. Solving the corresponding RGE at the one-loop order yields
\beq \label{eq:example1}
\frac{C_i (\mu)}{C_i (\mu_0)} \simeq \eta^{-\frac{g_i}{b}} \simeq 1 - \frac{1}{16 \pi^2} \hspace{0.25mm} \frac{\gamma_i (\mu_0)}{2} \hspace{0.25mm} \ln \left ( \frac{\mu_0^2}{\mu^2} \right ) \,,
\eeq
where we introduced 
\beq \label{eq:gi}
g_i = \frac{\gamma_i (\mu_0)}{\alpha_s (\mu_0)} \,, 
\eeq
with $\gamma_i (\mu_0)$ the scale-independent anomalous dimension describing the self-mixing of $Q_i$. Observe that in the final expression of~(\ref{eq:example1}), we have expanded the resummed result, retaining only the LL term. From~(\ref{eq:example1}), one can directly obtain the first line of~(\ref{eq:ratioC}).

The second type of RG evolution we consider involves the mixing of an operator $Q_i$ into $Q_j$, with both operators potentially having nonzero self-mixing. Assuming $C_j (\mu_0) = 0$ and that the anomalous dimension $\gamma_{ji}$, which governs this mixing, receives a one-loop contribution proportional to $g_s^2 \hspace{0.125mm} y_t$, we obtain
\beq \label{eq:example2}
\frac{C_j (\mu)}{C_i (\mu_0)} \simeq -\frac{\gamma_{ji} (\mu_0)}{\alpha_s (\mu)} \frac{1}{y - g_i + g_ j} \left ( \eta^{\frac{b + y - g_i}{b}} - \eta^{\frac{b - g_j}{b}} \right ) \simeq - \frac{1}{16 \pi^2} \hspace{0.25mm} \frac{\gamma_{ji} (\mu_0)}{2} \hspace{0.25mm} \ln \left ( \frac{\mu_0^2}{\mu^2} \right ) \,,
\eeq
where the symbols $b$, $y$, and $g_i$ are defined in~(\ref{eq:gendef}) and (\ref{eq:gi}), respectively. We emphasize~that in the first expression of~(\ref{eq:example2}), the logarithms associated with the factors~$g_s$ and~$y_t$ that~appear in all the anomalous dimensions are resummed. The final result, on the other hand, retains only the LL term, which is proportional to the scale-independent anomalous dimension $\gamma_{ji} (\mu_0)$ that describes the mixing of $Q_i$ into $Q_j$. Notice that~(\ref{eq:example2}) allows for a direct derivation of the second line in~(\ref{eq:ratioC}).

The third type of RG corrections we aim to resum involves two-step mixing processes. Here, the running occurs through the one-loop mixing of the operator $Q_i$ into an intermediate operator $Q_m$, which then undergoes one-loop mixing into $Q_j$. Assuming $C_j (\mu_0) = C_m (\mu_0) = 0$, and that $\gamma_{mi}$ and $\gamma_{jm}$ receive one-loop contributions proportional to $g_s^2 \hspace{0.125mm} y_t$ and $g_s \hspace{0.125mm} y_t$, respectively, while $\gamma_{ji} = 0$ at this order, we find 
\bea \label{eq:example3}
\begin{split}
\frac{C_j (\mu)}{C_i (\mu_0)} & \simeq -\frac{\gamma_{jm} (\mu_0) \hspace{0.25mm} \gamma_{mi} (\mu_0)}{\alpha_s^2 (\mu)} \left [ \, \frac{2}{y - g_i + g_m} \left ( \frac{\eta^{\frac{3 b + 4 y - 2 g_i}{2 b}}}{b - 4 y + 2 g_i - 2 g_j} - \frac{\eta^{\frac{3 b + 2 y - 2 g_m}{2 b}}}{b - 2 y - 2 g_j + 2 g_m} \right ) \right . \hspace{2mm} \\[2mm]
& \hspace{3.75cm} \left. - \frac{4 \hspace{0.125mm} \eta^{\frac{2 b - g_j}{b}}}{\left ( b - 4 y + 2 g_i - 2 g_j \right ) \left ( b - 2 y - 2 g_j + 2 g_m \right )}\, \right ] \\[4mm]
& \simeq \frac{1}{512 \pi^4} \hspace{0.25mm} \frac{\gamma_{jm} (\mu_0) \hspace{0.25mm} \gamma_{mi} (\mu_0)}{4} \hspace{0.25mm} \ln^2 \left ( \frac{\mu_0^2}{\mu^2} \right ) \,.
\end{split} 
\eea
The expressions for $b$, $y$, and $g_i$ are provided in~(\ref{eq:gendef}) and (\ref{eq:gi}). The first expression in~(\ref{eq:example3}) represents the resummed result, while the final expression corresponds to the LL~term, which is proportional to the product $\gamma_{jm} (\mu_0) \hspace{0.25mm} \gamma_{mi} (\mu_0)$ of scale-independent anomalous dimensions. Notably, (\ref{eq:example3}) in combination with (\ref{eq:example2}) allows for deriving the final expression in (\ref{eq:ratioC}). In the case of the two-loop mixing of $Q_G$ into $Q_{HG}$, which is proportional to $g_s^3 \hspace{0.125mm} y_t^2$, the expression~(\ref{eq:example2}) applies after the replacements $y \to b/2 + 2 y$ and $\gamma_{ji} (\mu_0) \to \beta_{HG}^{(2)} (\mu_0)/\big (16 \pi^2 \hspace{0.25mm} C_{G} (\mu_0) \big )$.

\end{appendix}

%\bibliographystyle{apsrev4-1}
%\bibliography{main}

\begin{thebibliography}{66}%
\makeatletter
\providecommand \@ifxundefined [1]{%
 \@ifx{#1\undefined}
}%
\providecommand \@ifnum [1]{%
 \ifnum #1\expandafter \@firstoftwo
 \else \expandafter \@secondoftwo
 \fi
}%
\providecommand \@ifx [1]{%
 \ifx #1\expandafter \@firstoftwo
 \else \expandafter \@secondoftwo
 \fi
}%
\providecommand \natexlab [1]{#1}%
\providecommand \enquote [1]{``#1''}%
\providecommand \bibnamefont [1]{#1}%
\providecommand \bibfnamefont [1]{#1}%
\providecommand \citenamefont [1]{#1}%
\providecommand \href@noop [0]{\@secondoftwo}%
\providecommand \href [0]{\begingroup \@sanitize@url \@href}%
\providecommand \@href[1]{\@@startlink{#1}\@@href}%
\providecommand \@@href[1]{\endgroup#1\@@endlink}%
\providecommand \@sanitize@url [0]{\catcode `\\12\catcode `\$12\catcode
 `\&12\catcode `\#12\catcode `\^12\catcode `\_12\catcode `\%12\relax}%
\providecommand \@@startlink[1]{}%
\providecommand \@@endlink[0]{}%
\providecommand \url [0]{\begingroup\@sanitize@url \@url }%
\providecommand \@url [1]{\endgroup\@href {#1}{\urlprefix }}%
\providecommand \urlprefix [0]{URL }%
\providecommand \Eprint [0]{\href }%
\providecommand \doibase [0]{http://dx.doi.org/}%
\providecommand \selectlanguage [0]{\@gobble}%
\providecommand \bibinfo [0]{\@secondoftwo}%
\providecommand \bibfield [0]{\@secondoftwo}%
\providecommand \translation [1]{[#1]}%
\providecommand \BibitemOpen [0]{}%
\providecommand \bibitemStop [0]{}%
\providecommand \bibitemNoStop [0]{.\EOS\space}%
\providecommand \EOS [0]{\spacefactor3000\relax}%
\providecommand\BibitemShut [1]{\csname bibitem#1\endcsname}%
\let\auto@bib@innerbib\@empty
%</preamble>
\bibitem [{\citenamefont {Buchm{\"u}ller}\ and\ \citenamefont
 {Wyler}(1986)}]{Buchmuller:1985jz}%
 \BibitemOpen
 \bibfield {author} {\bibinfo {author} {\bibfnamefont {W.}~\bibnamefont
 {Buchm{\"u}ller}}\ and\ \bibinfo {author} {\bibfnamefont {D.}~\bibnamefont
 {Wyler}},\ }\href {\doibase 10.1016/0550-3213(86)90262-2} {\bibfield
 {journal} {\bibinfo {journal} {Nucl. Phys. B}\ }\textbf {\bibinfo {volume}
 {268}},\ \bibinfo {pages} {621} (\bibinfo {year} {1986})}\BibitemShut
 {NoStop}%
\bibitem [{\citenamefont {Grzadkowski}\ \emph {et~al.}(2010)\citenamefont
 {Grzadkowski}, \citenamefont {Iskrzynski}, \citenamefont {Misiak},\ and\
 \citenamefont {Rosiek}}]{Grzadkowski:2010es}%
 \BibitemOpen
 \bibfield {author} {\bibinfo {author} {\bibfnamefont {B.}~\bibnamefont
 {Grzadkowski}}, \bibinfo {author} {\bibfnamefont {M.}~\bibnamefont
 {Iskrzynski}}, \bibinfo {author} {\bibfnamefont {M.}~\bibnamefont {Misiak}},
 \ and\ \bibinfo {author} {\bibfnamefont {J.}~\bibnamefont {Rosiek}},\ }\href
 {\doibase 10.1007/JHEP10(2010)085} {\bibfield {journal} {\bibinfo {journal}
 {JHEP}\ }\textbf {\bibinfo {volume} {10}},\ \bibinfo {pages} {085} (\bibinfo
 {year} {2010})},\ \Eprint {http://arxiv.org/abs/1008.4884} {arXiv:1008.4884
 [hep-ph]}\BibitemShut {NoStop}%
\bibitem [{\citenamefont {Brivio}\ and\ \citenamefont
 {Trott}(2019)}]{Brivio:2017vri}%
 \BibitemOpen
 \bibfield {author} {\bibinfo {author} {\bibfnamefont {I.}~\bibnamefont
 {Brivio}}\ and\ \bibinfo {author} {\bibfnamefont {M.}~\bibnamefont {Trott}},\
 }\href {\doibase 10.1016/j.physrep.2018.11.002} {\bibfield {journal}
 {\bibinfo {journal} {Phys. Rept.}\ }\textbf {\bibinfo {volume} {793}},\
 \bibinfo {pages} {1} (\bibinfo {year} {2019})},\ \Eprint
 {http://arxiv.org/abs/1706.08945} {arXiv:1706.08945 [hep-ph]}\BibitemShut
 {NoStop}%
\bibitem [{\citenamefont {Isidori}\ \emph {et~al.}(2024)\citenamefont
 {Isidori}, \citenamefont {Wilsch},\ and\ \citenamefont
 {Wyler}}]{Isidori:2023pyp}%
 \BibitemOpen
 \bibfield {author} {\bibinfo {author} {\bibfnamefont {G.}~\bibnamefont
 {Isidori}}, \bibinfo {author} {\bibfnamefont {F.}~\bibnamefont {Wilsch}}, \
 and\ \bibinfo {author} {\bibfnamefont {D.}~\bibnamefont {Wyler}},\ }\href
 {\doibase 10.1103/RevModPhys.96.015006} {\bibfield {journal} {\bibinfo
 {journal} {Rev. Mod. Phys.}\ }\textbf {\bibinfo {volume} {96}},\ \bibinfo
 {pages} {015006} (\bibinfo {year} {2024})},\ \Eprint
 {http://arxiv.org/abs/2303.16922} {arXiv:2303.16922 [hep-ph]}\BibitemShut
 {NoStop}%
\bibitem [{\citenamefont {Degrande}\ \emph {et~al.}(2021)\citenamefont
 {Degrande}, \citenamefont {Durieux}, \citenamefont {Maltoni}, \citenamefont
 {Mimasu}, \citenamefont {Vryonidou},\ and\ \citenamefont
 {Zhang}}]{Degrande:2020evl}%
 \BibitemOpen
 \bibfield {author} {\bibinfo {author} {\bibfnamefont {C.}~\bibnamefont
 {Degrande}}, \bibinfo {author} {\bibfnamefont {G.}~\bibnamefont {Durieux}},
 \bibinfo {author} {\bibfnamefont {F.}~\bibnamefont {Maltoni}}, \bibinfo
 {author} {\bibfnamefont {K.}~\bibnamefont {Mimasu}}, \bibinfo {author}
 {\bibfnamefont {E.}~\bibnamefont {Vryonidou}}, \ and\ \bibinfo {author}
 {\bibfnamefont {C.}~\bibnamefont {Zhang}},\ }\href {\doibase
 10.1103/PhysRevD.103.096024} {\bibfield {journal} {\bibinfo {journal}
 {Phys. Rev. D}\ }\textbf {\bibinfo {volume} {103}},\ \bibinfo {pages}
 {096024} (\bibinfo {year} {2021})},\ \Eprint
 {http://arxiv.org/abs/2008.11743} {arXiv:2008.11743 [hep-ph]}\BibitemShut
 {NoStop}%
\bibitem [{\citenamefont {Jenkins}\ \emph {et~al.}(2013)\citenamefont
 {Jenkins}, \citenamefont {Manohar},\ and\ \citenamefont
 {Trott}}]{Jenkins:2013zja}%
 \BibitemOpen
 \bibfield {author} {\bibinfo {author} {\bibfnamefont {E.~E.}\ \bibnamefont
 {Jenkins}}, \bibinfo {author} {\bibfnamefont {A.~V.}\ \bibnamefont
 {Manohar}}, \ and\ \bibinfo {author} {\bibfnamefont {M.}~\bibnamefont
 {Trott}},\ }\href {\doibase 10.1007/JHEP10(2013)087} {\bibfield {journal}
 {\bibinfo {journal} {JHEP}\ }\textbf {\bibinfo {volume} {10}},\ \bibinfo
 {pages} {087} (\bibinfo {year} {2013})},\ \Eprint
 {http://arxiv.org/abs/1308.2627} {arXiv:1308.2627 [hep-ph]}\BibitemShut
 {NoStop}%
\bibitem [{\citenamefont {Jenkins}\ \emph {et~al.}(2014)\citenamefont
 {Jenkins}, \citenamefont {Manohar},\ and\ \citenamefont
 {Trott}}]{Jenkins:2013wua}%
 \BibitemOpen
 \bibfield {author} {\bibinfo {author} {\bibfnamefont {E.~E.}\ \bibnamefont
 {Jenkins}}, \bibinfo {author} {\bibfnamefont {A.~V.}\ \bibnamefont
 {Manohar}}, \ and\ \bibinfo {author} {\bibfnamefont {M.}~\bibnamefont
 {Trott}},\ }\href {\doibase 10.1007/JHEP01(2014)035} {\bibfield {journal}
 {\bibinfo {journal} {JHEP}\ }\textbf {\bibinfo {volume} {01}},\ \bibinfo
 {pages} {035} (\bibinfo {year} {2014})},\ \Eprint
 {http://arxiv.org/abs/1310.4838} {arXiv:1310.4838 [hep-ph]}\BibitemShut
 {NoStop}%
\bibitem [{\citenamefont {Alonso}\ \emph {et~al.}(2014)\citenamefont {Alonso},
 \citenamefont {Jenkins}, \citenamefont {Manohar},\ and\ \citenamefont
 {Trott}}]{Alonso:2013hga}%
 \BibitemOpen
 \bibfield {author} {\bibinfo {author} {\bibfnamefont {R.}~\bibnamefont
 {Alonso}}, \bibinfo {author} {\bibfnamefont {E.~E.}\ \bibnamefont {Jenkins}},
 \bibinfo {author} {\bibfnamefont {A.~V.}\ \bibnamefont {Manohar}}, \ and\
 \bibinfo {author} {\bibfnamefont {M.}~\bibnamefont {Trott}},\ }\href
 {\doibase 10.1007/JHEP04(2014)159} {\bibfield {journal} {\bibinfo {journal}
 {JHEP}\ }\textbf {\bibinfo {volume} {04}},\ \bibinfo {pages} {159} (\bibinfo
 {year} {2014})},\ \Eprint {http://arxiv.org/abs/1312.2014} {arXiv:1312.2014
 [hep-ph]}\BibitemShut {NoStop}%
\bibitem [{\citenamefont {Haisch}\ \emph {et~al.}(2022)\citenamefont {Haisch},
 \citenamefont {Scott}, \citenamefont {Wiesemann}, \citenamefont
 {Zanderighi},\ and\ \citenamefont {Zanoli}}]{Haisch:2022nwz}%
 \BibitemOpen
 \bibfield {author} {\bibinfo {author} {\bibfnamefont {U.}~\bibnamefont
 {Haisch}}, \bibinfo {author} {\bibfnamefont {D.~J.}\ \bibnamefont {Scott}},
 \bibinfo {author} {\bibfnamefont {M.}~\bibnamefont {Wiesemann}}, \bibinfo
 {author} {\bibfnamefont {G.}~\bibnamefont {Zanderighi}}, \ and\ \bibinfo
 {author} {\bibfnamefont {S.}~\bibnamefont {Zanoli}},\ }\href {\doibase
 10.1007/JHEP07(2022)054} {\bibfield {journal} {\bibinfo {journal} {JHEP}\
 }\textbf {\bibinfo {volume} {07}},\ \bibinfo {pages} {054} (\bibinfo {year}
 {2022})},\ \Eprint {http://arxiv.org/abs/2204.00663} {arXiv:2204.00663
 [hep-ph]}\BibitemShut {NoStop}%
\bibitem [{\citenamefont {Di~Noi}\ \emph {et~al.}(2024)\citenamefont {Di~Noi},
 \citenamefont {Gr\"ober}, \citenamefont {Heinrich}, \citenamefont {Lang},\
 and\ \citenamefont {Vitti}}]{DiNoi:2023ygk}%
 \BibitemOpen
 \bibfield {author} {\bibinfo {author} {\bibfnamefont {S.}~\bibnamefont
 {Di~Noi}}, \bibinfo {author} {\bibfnamefont {R.}~\bibnamefont {Gr\"ober}},
 \bibinfo {author} {\bibfnamefont {G.}~\bibnamefont {Heinrich}}, \bibinfo
 {author} {\bibfnamefont {J.}~\bibnamefont {Lang}}, \ and\ \bibinfo {author}
 {\bibfnamefont {M.}~\bibnamefont {Vitti}},\ }\href {\doibase
 10.1103/PhysRevD.109.095024} {\bibfield {journal} {\bibinfo {journal}
 {Phys. Rev. D}\ }\textbf {\bibinfo {volume} {109}},\ \bibinfo {pages}
 {095024} (\bibinfo {year} {2024})},\ \Eprint
 {http://arxiv.org/abs/2310.18221} {arXiv:2310.18221 [hep-ph]}\BibitemShut
 {NoStop}%
\bibitem [{\citenamefont {Gauld}\ \emph {et~al.}(2024)\citenamefont {Gauld},
 \citenamefont {Haisch},\ and\ \citenamefont {Schnell}}]{Gauld:2023gtb}%
 \BibitemOpen
 \bibfield {author} {\bibinfo {author} {\bibfnamefont {R.}~\bibnamefont
 {Gauld}}, \bibinfo {author} {\bibfnamefont {U.}~\bibnamefont {Haisch}}, \
 and\ \bibinfo {author} {\bibfnamefont {L.}~\bibnamefont {Schnell}},\ }\href
 {\doibase 10.1007/JHEP01(2024)192} {\bibfield {journal} {\bibinfo {journal}
 {JHEP}\ }\textbf {\bibinfo {volume} {01}},\ \bibinfo {pages} {192} (\bibinfo
 {year} {2024})},\ \Eprint {http://arxiv.org/abs/2311.06107} {arXiv:2311.06107
 [hep-ph]}\BibitemShut {NoStop}%
\bibitem [{\citenamefont {Heinrich}\ and\ \citenamefont
 {Lang}(2024)}]{Heinrich:2023rsd}%
 \BibitemOpen
 \bibfield {author} {\bibinfo {author} {\bibfnamefont {G.}~\bibnamefont
 {Heinrich}}\ and\ \bibinfo {author} {\bibfnamefont {J.}~\bibnamefont
 {Lang}},\ }\href {\doibase 10.1007/JHEP05(2024)121} {\bibfield {journal}
 {\bibinfo {journal} {JHEP}\ }\textbf {\bibinfo {volume} {05}},\ \bibinfo
 {pages} {121} (\bibinfo {year} {2024})},\ \Eprint
 {http://arxiv.org/abs/2311.15004} {arXiv:2311.15004 [hep-ph]}\BibitemShut
 {NoStop}%
\bibitem [{\citenamefont {Di~Noi}\ \emph {et~al.}(2025)\citenamefont {Di~Noi},
 \citenamefont {Gr\"ober},\ and\ \citenamefont {Mandal}}]{DiNoi:2024ajj}%
 \BibitemOpen
 \bibfield {author} {\bibinfo {author} {\bibfnamefont {S.}~\bibnamefont
 {Di~Noi}}, \bibinfo {author} {\bibfnamefont {R.}~\bibnamefont {Gr\"ober}}, \
 and\ \bibinfo {author} {\bibfnamefont {M.~K.}\ \bibnamefont {Mandal}},\
 }\href {\doibase 10.1007/JHEP12(2024)220} {\bibfield {journal} {\bibinfo
 {journal} {JHEP}\ }\textbf {\bibinfo {volume} {12}},\ \bibinfo {pages} {220}
 (\bibinfo {year} {2025})},\ \Eprint {http://arxiv.org/abs/2408.03252}
 {arXiv:2408.03252 [hep-ph]}\BibitemShut {NoStop}%
\bibitem [{\citenamefont {Born}\ \emph {et~al.}(2024)\citenamefont {Born},
 \citenamefont {Fuentes-Mart\'\i{}n}, \citenamefont {Kvedarait\.{e}},\ and\
 \citenamefont {Thomsen}}]{Born:2024mgz}%
 \BibitemOpen
 \bibfield {author} {\bibinfo {author} {\bibfnamefont {L.}~\bibnamefont
 {Born}}, \bibinfo {author} {\bibfnamefont {J.}~\bibnamefont
 {Fuentes-Mart\'\i{}n}}, \bibinfo {author} {\bibfnamefont {S.}~\bibnamefont
 {Kvedarait\.{e}}}, \ and\ \bibinfo {author} {\bibfnamefont {A.~E.}\
 \bibnamefont {Thomsen}},\ }\href@noop {} {\ (\bibinfo {year} {2024})},\
 \Eprint {http://arxiv.org/abs/2410.07320} {arXiv:2410.07320 [hep-ph]}
 \BibitemShut {NoStop}%
\bibitem [{\citenamefont {Haisch}\ and\ \citenamefont
 {Schnell}(2025)}]{Haisch:2024wnw}%
 \BibitemOpen
 \bibfield {author} {\bibinfo {author} {\bibfnamefont {U.}~\bibnamefont
 {Haisch}}\ and\ \bibinfo {author} {\bibfnamefont {L.}~\bibnamefont
 {Schnell}},\ }\href {\doibase 10.1007/JHEP02(2025)038} {\bibfield {journal}
 {\bibinfo {journal} {JHEP}\ }\textbf {\bibinfo {volume} {02}},\ \bibinfo
 {pages} {038} (\bibinfo {year} {2025})},\ \Eprint
 {http://arxiv.org/abs/2410.13304} {arXiv:2410.13304 [hep-ph]}\BibitemShut
 {NoStop}%
\bibitem [{\citenamefont {Bonetti}\ \emph {et~al.}(2025)\citenamefont
 {Bonetti}, \citenamefont {Harlander}, \citenamefont {Korneev}, \citenamefont
 {Long}, \citenamefont {Melnikov}, \citenamefont {R\"ontsch},\ and\
 \citenamefont {Tagliabue}}]{Bonetti:2025hnb}%
 \BibitemOpen
 \bibfield {author} {\bibinfo {author} {\bibfnamefont {M.}~\bibnamefont
 {Bonetti}}, \bibinfo {author} {\bibfnamefont {R.~V.}\ \bibnamefont
 {Harlander}}, \bibinfo {author} {\bibfnamefont {D.}~\bibnamefont {Korneev}},
 \bibinfo {author} {\bibfnamefont {M.-M.}\ \bibnamefont {Long}}, \bibinfo
 {author} {\bibfnamefont {K.}~\bibnamefont {Melnikov}}, \bibinfo {author}
 {\bibfnamefont {R.}~\bibnamefont {R\"ontsch}}, \ and\ \bibinfo {author}
 {\bibfnamefont {D.~M.}\ \bibnamefont {Tagliabue}},\ }\href@noop {} {\
 (\bibinfo {year} {2025})},\ \Eprint {http://arxiv.org/abs/2502.12846}
 {arXiv:2502.12846 [hep-ph]}\BibitemShut {NoStop}%
\bibitem [{\citenamefont {Ghosh}\ and\ \citenamefont
 {Wiebusch}(2015)}]{Ghosh:2014wxa}%
 \BibitemOpen
 \bibfield {author} {\bibinfo {author} {\bibfnamefont {D.}~\bibnamefont
 {Ghosh}}\ and\ \bibinfo {author} {\bibfnamefont {M.}~\bibnamefont
 {Wiebusch}},\ }\href {\doibase 10.1103/PhysRevD.91.031701} {\bibfield
 {journal} {\bibinfo {journal} {Phys. Rev. D}\ }\textbf {\bibinfo {volume}
 {91}},\ \bibinfo {pages} {031701} (\bibinfo {year} {2015})},\ \Eprint
 {http://arxiv.org/abs/1411.2029} {arXiv:1411.2029 [hep-ph]}\BibitemShut
 {NoStop}%
\bibitem [{\citenamefont {Krauss}\ \emph {et~al.}(2017)\citenamefont {Krauss},
 \citenamefont {Kuttimalai},\ and\ \citenamefont {Plehn}}]{Krauss:2016ely}%
 \BibitemOpen
 \bibfield {author} {\bibinfo {author} {\bibfnamefont {F.}~\bibnamefont
 {Krauss}}, \bibinfo {author} {\bibfnamefont {S.}~\bibnamefont {Kuttimalai}},
 \ and\ \bibinfo {author} {\bibfnamefont {T.}~\bibnamefont {Plehn}},\ }\href
 {\doibase 10.1103/PhysRevD.95.035024} {\bibfield {journal} {\bibinfo
 {journal} {Phys. Rev. D}\ }\textbf {\bibinfo {volume} {95}},\ \bibinfo
 {pages} {035024} (\bibinfo {year} {2017})},\ \Eprint
 {http://arxiv.org/abs/1611.00767} {arXiv:1611.00767 [hep-ph]}\BibitemShut
 {NoStop}%
\bibitem [{\citenamefont {Hirschi}\ \emph {et~al.}(2018)\citenamefont
 {Hirschi}, \citenamefont {Maltoni}, \citenamefont {Tsinikos},\ and\
 \citenamefont {Vryonidou}}]{Hirschi:2018etq}%
 \BibitemOpen
 \bibfield {author} {\bibinfo {author} {\bibfnamefont {V.}~\bibnamefont
 {Hirschi}}, \bibinfo {author} {\bibfnamefont {F.}~\bibnamefont {Maltoni}},
 \bibinfo {author} {\bibfnamefont {I.}~\bibnamefont {Tsinikos}}, \ and\
 \bibinfo {author} {\bibfnamefont {E.}~\bibnamefont {Vryonidou}},\ }\href
 {\doibase 10.1007/JHEP07(2018)093} {\bibfield {journal} {\bibinfo {journal}
 {JHEP}\ }\textbf {\bibinfo {volume} {07}},\ \bibinfo {pages} {093} (\bibinfo
 {year} {2018})},\ \Eprint {http://arxiv.org/abs/1806.04696} {arXiv:1806.04696
 [hep-ph]}\BibitemShut {NoStop}%
\bibitem [{\citenamefont {Goldouzian}\ and\ \citenamefont
 {Hildreth}(2020)}]{Goldouzian:2020wdq}%
 \BibitemOpen
 \bibfield {author} {\bibinfo {author} {\bibfnamefont {R.}~\bibnamefont
 {Goldouzian}}\ and\ \bibinfo {author} {\bibfnamefont {M.~D.}\ \bibnamefont
 {Hildreth}},\ }\href {\doibase 10.1016/j.physletb.2020.135889} {\bibfield
 {journal} {\bibinfo {journal} {Phys. Lett. B}\ }\textbf {\bibinfo {volume}
 {811}},\ \bibinfo {pages} {135889} (\bibinfo {year} {2020})},\ \Eprint
 {http://arxiv.org/abs/2001.02736} {arXiv:2001.02736 [hep-ph]}\BibitemShut
 {NoStop}%
\bibitem [{\citenamefont {Bardhan}\ \emph {et~al.}(2021)\citenamefont
 {Bardhan}, \citenamefont {Ghosh}, \citenamefont {Jain},\ and\ \citenamefont
 {Thalapillil}}]{Bardhan:2020vcl}%
 \BibitemOpen
 \bibfield {author} {\bibinfo {author} {\bibfnamefont {D.}~\bibnamefont
 {Bardhan}}, \bibinfo {author} {\bibfnamefont {D.}~\bibnamefont {Ghosh}},
 \bibinfo {author} {\bibfnamefont {P.}~\bibnamefont {Jain}}, \ and\ \bibinfo
 {author} {\bibfnamefont {A.~M.}\ \bibnamefont {Thalapillil}},\ }\href
 {\doibase 10.1103/PhysRevD.103.115003} {\bibfield {journal} {\bibinfo
 {journal} {Phys. Rev. D}\ }\textbf {\bibinfo {volume} {103}},\ \bibinfo
 {pages} {115003} (\bibinfo {year} {2021})},\ \Eprint
 {http://arxiv.org/abs/2010.13402} {arXiv:2010.13402 [hep-ph]}\BibitemShut
 {NoStop}%
\bibitem [{\citenamefont {Ellis}\ \emph {et~al.}(2021)\citenamefont {Ellis},
 \citenamefont {Madigan}, \citenamefont {Mimasu}, \citenamefont {Sanz},\ and\
 \citenamefont {You}}]{Ellis:2020unq}%
 \BibitemOpen
 \bibfield {author} {\bibinfo {author} {\bibfnamefont {J.}~\bibnamefont
 {Ellis}}, \bibinfo {author} {\bibfnamefont {M.}~\bibnamefont {Madigan}},
 \bibinfo {author} {\bibfnamefont {K.}~\bibnamefont {Mimasu}}, \bibinfo
 {author} {\bibfnamefont {V.}~\bibnamefont {Sanz}}, \ and\ \bibinfo {author}
 {\bibfnamefont {T.}~\bibnamefont {You}},\ }\href {\doibase
 10.1007/JHEP04(2021)279} {\bibfield {journal} {\bibinfo {journal} {JHEP}\
 }\textbf {\bibinfo {volume} {04}},\ \bibinfo {pages} {279} (\bibinfo {year}
 {2021})},\ \Eprint {http://arxiv.org/abs/2012.02779} {arXiv:2012.02779
 [hep-ph]}\BibitemShut {NoStop}%
\bibitem [{\citenamefont {Chanowitz}\ \emph {et~al.}(1979)\citenamefont
 {Chanowitz}, \citenamefont {Furman},\ and\ \citenamefont
 {Hinchliffe}}]{Chanowitz:1979zu}%
 \BibitemOpen
 \bibfield {author} {\bibinfo {author} {\bibfnamefont {M.~S.}\ \bibnamefont
 {Chanowitz}}, \bibinfo {author} {\bibfnamefont {M.}~\bibnamefont {Furman}}, \
 and\ \bibinfo {author} {\bibfnamefont {I.}~\bibnamefont {Hinchliffe}},\
 }\href {\doibase 10.1016/0550-3213(79)90333-X} {\bibfield {journal}
 {\bibinfo {journal} {Nucl. Phys. B}\ }\textbf {\bibinfo {volume} {159}},\
 \bibinfo {pages} {225} (\bibinfo {year} {1979})}\BibitemShut {NoStop}%
\bibitem [{\citenamefont {Alloul}\ \emph {et~al.}(2014)\citenamefont {Alloul},
 \citenamefont {Christensen}, \citenamefont {Degrande}, \citenamefont {Duhr},\
 and\ \citenamefont {Fuks}}]{Alloul:2013bka}%
 \BibitemOpen
 \bibfield {author} {\bibinfo {author} {\bibfnamefont {A.}~\bibnamefont
 {Alloul}}, \bibinfo {author} {\bibfnamefont {N.~D.}\ \bibnamefont
 {Christensen}}, \bibinfo {author} {\bibfnamefont {C.}~\bibnamefont
 {Degrande}}, \bibinfo {author} {\bibfnamefont {C.}~\bibnamefont {Duhr}}, \
 and\ \bibinfo {author} {\bibfnamefont {B.}~\bibnamefont {Fuks}},\ }\href
 {\doibase 10.1016/j.cpc.2014.04.012} {\bibfield {journal} {\bibinfo
 {journal} {Comput. Phys. Commun.}\ }\textbf {\bibinfo {volume} {185}},\
 \bibinfo {pages} {2250} (\bibinfo {year} {2014})},\ \Eprint
 {http://arxiv.org/abs/1310.1921} {arXiv:1310.1921 [hep-ph]}\BibitemShut
 {NoStop}%
\bibitem [{\citenamefont {Hahn}(2001)}]{Hahn:2000kx}%
 \BibitemOpen
 \bibfield {author} {\bibinfo {author} {\bibfnamefont {T.}~\bibnamefont
 {Hahn}},\ }\href {\doibase 10.1016/S0010-4655(01)00290-9} {\bibfield
 {journal} {\bibinfo {journal} {Comput. Phys. Commun.}\ }\textbf {\bibinfo
 {volume} {140}},\ \bibinfo {pages} {418} (\bibinfo {year} {2001})},\ \Eprint
 {http://arxiv.org/abs/hep-ph/0012260} {arXiv:hep-ph/0012260}\BibitemShut
 {NoStop}%
\bibitem [{\citenamefont {Hahn}\ \emph {et~al.}(2016)\citenamefont {Hahn},
 \citenamefont {Pa\ss{}ehr},\ and\ \citenamefont
 {Schappacher}}]{Hahn:2016ebn}%
 \BibitemOpen
 \bibfield {author} {\bibinfo {author} {\bibfnamefont {T.}~\bibnamefont
 {Hahn}}, \bibinfo {author} {\bibfnamefont {S.}~\bibnamefont {Pa\ss{}ehr}}, \
 and\ \bibinfo {author} {\bibfnamefont {C.}~\bibnamefont {Schappacher}},\
 }\href {\doibase 10.1088/1742-6596/762/1/012065} {\bibfield {journal}
 {\bibinfo {journal} {PoS}\ }\textbf {\bibinfo {volume} {LL2016}},\ \bibinfo
 {pages} {068} (\bibinfo {year} {2016})},\ \Eprint
 {http://arxiv.org/abs/1604.04611} {arXiv:1604.04611 [hep-ph]}\BibitemShut
 {NoStop}%
\bibitem [{\citenamefont {Lee}(2014)}]{Lee:2013mka}%
 \BibitemOpen
 \bibfield {author} {\bibinfo {author} {\bibfnamefont {R.~N.}\ \bibnamefont
 {Lee}},\ }\href {\doibase 10.1088/1742-6596/523/1/012059} {\bibfield
 {journal} {\bibinfo {journal} {J. Phys. Conf. Ser.}\ }\textbf {\bibinfo
 {volume} {523}},\ \bibinfo {pages} {012059} (\bibinfo {year} {2014})},\
 \Eprint {http://arxiv.org/abs/1310.1145} {arXiv:1310.1145 [hep-ph]}
 \BibitemShut {NoStop}%
\bibitem [{\citenamefont {Steinhauser}(2002)}]{Steinhauser:2002rq}%
 \BibitemOpen
 \bibfield {author} {\bibinfo {author} {\bibfnamefont {M.}~\bibnamefont
 {Steinhauser}},\ }\href {\doibase 10.1016/S0370-1573(02)00017-0} {\bibfield
 {journal} {\bibinfo {journal} {Phys. Rept.}\ }\textbf {\bibinfo {volume}
 {364}},\ \bibinfo {pages} {247} (\bibinfo {year} {2002})},\ \Eprint
 {http://arxiv.org/abs/hep-ph/0201075} {arXiv:hep-ph/0201075}\BibitemShut
 {NoStop}%
\bibitem [{\citenamefont {Anastasiou}\ \emph {et~al.}(2007)\citenamefont
 {Anastasiou}, \citenamefont {Beerli}, \citenamefont {Bucherer}, \citenamefont
 {Daleo},\ and\ \citenamefont {Kunszt}}]{Anastasiou:2006hc}%
 \BibitemOpen
 \bibfield {author} {\bibinfo {author} {\bibfnamefont {C.}~\bibnamefont
 {Anastasiou}}, \bibinfo {author} {\bibfnamefont {S.}~\bibnamefont {Beerli}},
 \bibinfo {author} {\bibfnamefont {S.}~\bibnamefont {Bucherer}}, \bibinfo
 {author} {\bibfnamefont {A.}~\bibnamefont {Daleo}}, \ and\ \bibinfo {author}
 {\bibfnamefont {Z.}~\bibnamefont {Kunszt}},\ }\href {\doibase
 10.1088/1126-6708/2007/01/082} {\bibfield {journal} {\bibinfo {journal}
 {JHEP}\ }\textbf {\bibinfo {volume} {01}},\ \bibinfo {pages} {082} (\bibinfo
 {year} {2007})},\ \Eprint {http://arxiv.org/abs/hep-ph/0611236}
 {arXiv:hep-ph/0611236}\BibitemShut {NoStop}%
\bibitem [{\citenamefont {Kotikov}(1991)}]{Kotikov:1990kg}%
 \BibitemOpen
 \bibfield {author} {\bibinfo {author} {\bibfnamefont {A.~V.}\ \bibnamefont
 {Kotikov}},\ }\href {\doibase 10.1016/0370-2693(91)90413-K} {\bibfield
 {journal} {\bibinfo {journal} {Phys. Lett. B}\ }\textbf {\bibinfo {volume}
 {254}},\ \bibinfo {pages} {158} (\bibinfo {year} {1991})}\BibitemShut
 {NoStop}%
\bibitem [{\citenamefont {Remiddi}(1997)}]{Remiddi:1997ny}%
 \BibitemOpen
 \bibfield {author} {\bibinfo {author} {\bibfnamefont {E.}~\bibnamefont
 {Remiddi}},\ }\href {\doibase 10.1007/BF03185566} {\bibfield {journal}
 {\bibinfo {journal} {Nuovo Cim. A}\ }\textbf {\bibinfo {volume} {110}},\
 \bibinfo {pages} {1435} (\bibinfo {year} {1997})},\ \Eprint
 {http://arxiv.org/abs/hep-th/9711188} {arXiv:hep-th/9711188}\BibitemShut
 {NoStop}%
\bibitem [{\citenamefont {Gehrmann}\ and\ \citenamefont
 {Remiddi}(2000)}]{Gehrmann:1999as}%
 \BibitemOpen
 \bibfield {author} {\bibinfo {author} {\bibfnamefont {T.}~\bibnamefont
 {Gehrmann}}\ and\ \bibinfo {author} {\bibfnamefont {E.}~\bibnamefont
 {Remiddi}},\ }\href {\doibase 10.1016/S0550-3213(00)00223-6} {\bibfield
 {journal} {\bibinfo {journal} {Nucl. Phys. B}\ }\textbf {\bibinfo {volume}
 {580}},\ \bibinfo {pages} {485} (\bibinfo {year} {2000})},\ \Eprint
 {http://arxiv.org/abs/hep-ph/9912329} {arXiv:hep-ph/9912329}\BibitemShut
 {NoStop}%
\bibitem [{\citenamefont {Argeri}\ and\ \citenamefont
 {Mastrolia}(2007)}]{Argeri:2007upc}%
 \BibitemOpen
 \bibfield {author} {\bibinfo {author} {\bibfnamefont {M.}~\bibnamefont
 {Argeri}}\ and\ \bibinfo {author} {\bibfnamefont {P.}~\bibnamefont
 {Mastrolia}},\ }\href {\doibase 10.1142/S0217751X07037147} {\bibfield
 {journal} {\bibinfo {journal} {Int. J. Mod. Phys. A}\ }\textbf {\bibinfo
 {volume} {22}},\ \bibinfo {pages} {4375} (\bibinfo {year} {2007})},\ \Eprint
 {http://arxiv.org/abs/0707.4037} {arXiv:0707.4037 [hep-ph]}\BibitemShut
 {NoStop}%
\bibitem [{\citenamefont {Henn}(2015)}]{Henn:2014qga}%
 \BibitemOpen
 \bibfield {author} {\bibinfo {author} {\bibfnamefont {J.~M.}\ \bibnamefont
 {Henn}},\ }\href {\doibase 10.1088/1751-8113/48/15/153001} {\bibfield
 {journal} {\bibinfo {journal} {J. Phys. A}\ }\textbf {\bibinfo {volume}
 {48}},\ \bibinfo {pages} {153001} (\bibinfo {year} {2015})},\ \Eprint
 {http://arxiv.org/abs/1412.2296} {arXiv:1412.2296 [hep-ph]}\BibitemShut
 {NoStop}%
\bibitem [{\citenamefont {Liu}\ and\ \citenamefont {Ma}(2023)}]{Liu:2022chg}%
 \BibitemOpen
 \bibfield {author} {\bibinfo {author} {\bibfnamefont {X.}~\bibnamefont
 {Liu}}\ and\ \bibinfo {author} {\bibfnamefont {Y.-Q.}\ \bibnamefont {Ma}},\
 }\href {\doibase 10.1016/j.cpc.2022.108565} {\bibfield {journal} {\bibinfo
 {journal} {Comput. Phys. Commun.}\ }\textbf {\bibinfo {volume} {283}},\
 \bibinfo {pages} {108565} (\bibinfo {year} {2023})},\ \Eprint
 {http://arxiv.org/abs/2201.11669} {arXiv:2201.11669 [hep-ph]}\BibitemShut
 {NoStop}%
 \bibitem [{\citenamefont {Bardeen}\ \emph {et~al.}(1978)\citenamefont
 {Bardeen}, \citenamefont {Buras}, \citenamefont {Duke},\ and\ \citenamefont
 {Muta}}]{Bardeen:1978yd}%
\BibitemOpen
 \bibfield {author} {\bibinfo {author} {\bibfnamefont {W.~A.}\ \bibnamefont
 {Bardeen}}, \bibinfo {author} {\bibfnamefont {A.~J.}\ \bibnamefont {Buras}},
 \bibinfo {author} {\bibfnamefont {D.~W.}\ \bibnamefont {Duke}}, \ and\
 \bibinfo {author} {\bibfnamefont {T.}~\bibnamefont {Muta}},\ }\href {\doibase
 10.1103/PhysRevD.18.3998} {\bibfield {journal} {\bibinfo {journal} {Phys.
 Rev. D}\ }\textbf {\bibinfo {volume} {18}},\ \bibinfo {pages} {3998}
 (\bibinfo {year} {1978})}\BibitemShut {NoStop}%
\bibitem [{\citenamefont {Chetyrkin}\ \emph {et~al.}(1998)\citenamefont
 {Chetyrkin}, \citenamefont {Misiak},\ and\ \citenamefont
 {M{\"u}nz}}]{Chetyrkin:1997fm}%
 \BibitemOpen
 \bibfield {author} {\bibinfo {author} {\bibfnamefont {K.~G.}\ \bibnamefont
 {Chetyrkin}}, \bibinfo {author} {\bibfnamefont {M.}~\bibnamefont {Misiak}}, \
 and\ \bibinfo {author} {\bibfnamefont {M.}~\bibnamefont {M{\"u}nz}},\ }\href
 {\doibase 10.1016/S0550-3213(98)00122-9} {\bibfield {journal} {\bibinfo
 {journal} {Nucl. Phys. B}\ }\textbf {\bibinfo {volume} {518}},\ \bibinfo
 {pages} {473} (\bibinfo {year} {1998})},\ \Eprint
 {http://arxiv.org/abs/hep-ph/9711266} {arXiv:hep-ph/9711266}\BibitemShut
 {NoStop}%
\bibitem [{\citenamefont {Gambino}\ \emph {et~al.}(2003)\citenamefont
 {Gambino}, \citenamefont {Gorbahn},\ and\ \citenamefont
 {Haisch}}]{Gambino:2003zm}%
 \BibitemOpen
 \bibfield {author} {\bibinfo {author} {\bibfnamefont {P.}~\bibnamefont
 {Gambino}}, \bibinfo {author} {\bibfnamefont {M.}~\bibnamefont {Gorbahn}}, \
 and\ \bibinfo {author} {\bibfnamefont {U.}~\bibnamefont {Haisch}},\ }\href
 {\doibase 10.1016/j.nuclphysb.2003.09.024} {\bibfield {journal} {\bibinfo
 {journal} {Nucl. Phys. B}\ }\textbf {\bibinfo {volume} {673}},\ \bibinfo
 {pages} {238} (\bibinfo {year} {2003})},\ \Eprint
 {http://arxiv.org/abs/hep-ph/0306079} {arXiv:hep-ph/0306079}\BibitemShut
 {NoStop}%
\bibitem [{\citenamefont {Gorbahn}\ and\ \citenamefont
 {Haisch}(2016)}]{Gorbahn:2016uoy}%
 \BibitemOpen
 \bibfield {author} {\bibinfo {author} {\bibfnamefont {M.}~\bibnamefont
 {Gorbahn}}\ and\ \bibinfo {author} {\bibfnamefont {U.}~\bibnamefont
 {Haisch}},\ }\href {\doibase 10.1007/JHEP10(2016)094} {\bibfield {journal}
 {\bibinfo {journal} {JHEP}\ }\textbf {\bibinfo {volume} {10}},\ \bibinfo
 {pages} {094} (\bibinfo {year} {2016})},\ \Eprint
 {http://arxiv.org/abs/1607.03773} {arXiv:1607.03773 [hep-ph]}\BibitemShut
 {NoStop}%
\bibitem [{\citenamefont {Bern}\ \emph {et~al.}(2020)\citenamefont {Bern},
 \citenamefont {Parra-Martinez},\ and\ \citenamefont {Sawyer}}]{Bern:2020ikv}%
 \BibitemOpen
 \bibfield {author} {\bibinfo {author} {\bibfnamefont {Z.}~\bibnamefont
 {Bern}}, \bibinfo {author} {\bibfnamefont {J.}~\bibnamefont
 {Parra-Martinez}}, \ and\ \bibinfo {author} {\bibfnamefont {E.}~\bibnamefont
 {Sawyer}},\ }\href {\doibase 10.1007/JHEP10(2020)211} {\bibfield {journal}
 {\bibinfo {journal} {JHEP}\ }\textbf {\bibinfo {volume} {10}},\ \bibinfo
 {pages} {211} (\bibinfo {year} {2020})},\ \Eprint
 {http://arxiv.org/abs/2005.12917} {arXiv:2005.12917 [hep-ph]}\BibitemShut
 {NoStop}%
\bibitem [{\citenamefont {Jin}\ \emph {et~al.}(2021)\citenamefont {Jin},
 \citenamefont {Ren},\ and\ \citenamefont {Yang}}]{Jin:2020pwh}%
 \BibitemOpen
 \bibfield {author} {\bibinfo {author} {\bibfnamefont {Q.}~\bibnamefont
 {Jin}}, \bibinfo {author} {\bibfnamefont {K.}~\bibnamefont {Ren}}, \ and\
 \bibinfo {author} {\bibfnamefont {G.}~\bibnamefont {Yang}},\ }\href {\doibase
 10.1007/JHEP04(2021)180} {\bibfield {journal} {\bibinfo {journal} {JHEP}\
 }\textbf {\bibinfo {volume} {04}},\ \bibinfo {pages} {180} (\bibinfo {year}
 {2021})},\ \Eprint {http://arxiv.org/abs/2011.02494} {arXiv:2011.02494
 [hep-ph]}\BibitemShut {NoStop}%
\bibitem [{\citenamefont {Jenkins}\ \emph {et~al.}(2024)\citenamefont
 {Jenkins}, \citenamefont {Manohar}, \citenamefont {Naterop},\ and\
 \citenamefont {Pag\`es}}]{Jenkins:2023bls}%
 \BibitemOpen
 \bibfield {author} {\bibinfo {author} {\bibfnamefont {E.~E.}\ \bibnamefont
 {Jenkins}}, \bibinfo {author} {\bibfnamefont {A.~V.}\ \bibnamefont
 {Manohar}}, \bibinfo {author} {\bibfnamefont {L.}~\bibnamefont {Naterop}}, \
 and\ \bibinfo {author} {\bibfnamefont {J.}~\bibnamefont {Pag\`es}},\ }\href
 {\doibase 10.1007/JHEP02(2024)131} {\bibfield {journal} {\bibinfo {journal}
 {JHEP}\ }\textbf {\bibinfo {volume} {02}},\ \bibinfo {pages} {131} (\bibinfo
 {year} {2024})},\ \Eprint {http://arxiv.org/abs/2310.19883} {arXiv:2310.19883
 [hep-ph]}\BibitemShut {NoStop}%
 \bibitem [{\citenamefont {Duhr}\ \emph {et~al.}(2025)\citenamefont {Duhr},
 \citenamefont {Vasquez}, \citenamefont {Ventura},\ and\ \citenamefont
 {Vryonidou}}]{Duhr:2025zqw}%
 \BibitemOpen
 \bibfield {author} {\bibinfo {author} {\bibfnamefont {C.}~\bibnamefont
 {Duhr}}, \bibinfo {author} {\bibfnamefont {A.}~\bibnamefont {Vasquez}},
 \bibinfo {author} {\bibfnamefont {G.}~\bibnamefont {Ventura}}, \ and\
 \bibinfo {author} {\bibfnamefont {E.}~\bibnamefont {Vryonidou}},\ }\href@noop
 {} {\ (\bibinfo {year} {2025})},\ \Eprint {http://arxiv.org/abs/2503.01954}
 {arXiv:2503.01954 [hep-ph]}\BibitemShut {NoStop}%
\bibitem [{\citenamefont {Born}\ \emph {et~al.}(2026)\citenamefont {Born},
  \citenamefont {Fuentes-Mart{\'\i}n},\ and\ \citenamefont
  {Thomsen}}]{Born:2026xkr}%
  \BibitemOpen
  \bibfield  {author} {\bibinfo {author} {\bibfnamefont {L.}~\bibnamefont
  {Born}}, \bibinfo {author} {\bibfnamefont {J.}~\bibnamefont
  {Fuentes-Mart{\'\i}n}}, \ and\ \bibinfo {author} {\bibfnamefont {A.~E.}\
  \bibnamefont {Thomsen}},\ }\href@noop {} {\  (\bibinfo {year} {2026})},\
  \Eprint {http://arxiv.org/abs/2601.19974} {arXiv:2601.19974 [hep-ph]}\BibitemShut {NoStop}% 
\bibitem [{\citenamefont {Grazzini}\ \emph {et~al.}(2017)\citenamefont
 {Grazzini}, \citenamefont {Ilnicka}, \citenamefont {Spira},\ and\
 \citenamefont {Wiesemann}}]{Grazzini:2016paz}%
 \BibitemOpen
 \bibfield {author} {\bibinfo {author} {\bibfnamefont {M.}~\bibnamefont
 {Grazzini}}, \bibinfo {author} {\bibfnamefont {A.}~\bibnamefont {Ilnicka}},
 \bibinfo {author} {\bibfnamefont {M.}~\bibnamefont {Spira}}, \ and\ \bibinfo
 {author} {\bibfnamefont {M.}~\bibnamefont {Wiesemann}},\ }\href {\doibase
 10.1007/JHEP03(2017)115} {\bibfield {journal} {\bibinfo {journal} {JHEP}\
 }\textbf {\bibinfo {volume} {03}},\ \bibinfo {pages} {115} (\bibinfo {year}
 {2017})},\ \Eprint {http://arxiv.org/abs/1612.00283} {arXiv:1612.00283
 [hep-ph]}\BibitemShut {NoStop}%
\bibitem [{\citenamefont {Deutschmann}\ \emph {et~al.}(2017)\citenamefont
 {Deutschmann}, \citenamefont {Duhr}, \citenamefont {Maltoni},\ and\
 \citenamefont {Vryonidou}}]{Deutschmann:2017qum}%
 \BibitemOpen
 \bibfield {author} {\bibinfo {author} {\bibfnamefont {N.}~\bibnamefont
 {Deutschmann}}, \bibinfo {author} {\bibfnamefont {C.}~\bibnamefont {Duhr}},
 \bibinfo {author} {\bibfnamefont {F.}~\bibnamefont {Maltoni}}, \ and\
 \bibinfo {author} {\bibfnamefont {E.}~\bibnamefont {Vryonidou}},\ }\href
 {\doibase 10.1007/JHEP12(2017)063} {\bibfield {journal} {\bibinfo {journal}
 {JHEP}\ }\textbf {\bibinfo {volume} {12}},\ \bibinfo {pages} {063} (\bibinfo
 {year} {2017})},\ \bibinfo {note} {[Erratum: JHEP {\bf 02}, 159 (2018)]},\ \Eprint
 {http://arxiv.org/abs/1708.00460} {arXiv:1708.00460 [hep-ph]}\BibitemShut
 {NoStop}%
\bibitem [{\citenamefont {Grazzini}\ \emph {et~al.}(2018)\citenamefont
 {Grazzini}, \citenamefont {Ilnicka},\ and\ \citenamefont
 {Spira}}]{Grazzini:2018eyk}%
 \BibitemOpen
 \bibfield {author} {\bibinfo {author} {\bibfnamefont {M.}~\bibnamefont
 {Grazzini}}, \bibinfo {author} {\bibfnamefont {A.}~\bibnamefont {Ilnicka}}, \
 and\ \bibinfo {author} {\bibfnamefont {M.}~\bibnamefont {Spira}},\ }\href
 {\doibase 10.1140/epjc/s10052-018-6261-7} {\bibfield {journal} {\bibinfo
 {journal} {Eur. Phys. J. C}\ }\textbf {\bibinfo {volume} {78}},\ \bibinfo
 {pages} {808} (\bibinfo {year} {2018})},\ \Eprint
 {http://arxiv.org/abs/1806.08832} {arXiv:1806.08832 [hep-ph]}\BibitemShut
 {NoStop}%
\bibitem [{\citenamefont {Hisano}\ \emph {et~al.}(2012)\citenamefont {Hisano},
 \citenamefont {Tsumura},\ and\ \citenamefont {Yang}}]{Hisano:2012cc}%
 \BibitemOpen
 \bibfield {author} {\bibinfo {author} {\bibfnamefont {J.}~\bibnamefont
 {Hisano}}, \bibinfo {author} {\bibfnamefont {K.}~\bibnamefont {Tsumura}}, \
 and\ \bibinfo {author} {\bibfnamefont {M.~J.~S.}\ \bibnamefont {Yang}},\
 }\href {\doibase 10.1016/j.physletb.2012.06.038} {\bibfield {journal}
 {\bibinfo {journal} {Phys. Lett. B}\ }\textbf {\bibinfo {volume} {713}},\
 \bibinfo {pages} {473} (\bibinfo {year} {2012})},\ \Eprint
 {http://arxiv.org/abs/1205.2212} {arXiv:1205.2212 [hep-ph]}\BibitemShut
 {NoStop}%
\bibitem [{\citenamefont {Brod}\ \emph {et~al.}(2013)\citenamefont {Brod},
 \citenamefont {Haisch},\ and\ \citenamefont {Zupan}}]{Brod:2013cka}%
 \BibitemOpen
 \bibfield {author} {\bibinfo {author} {\bibfnamefont {J.}~\bibnamefont
 {Brod}}, \bibinfo {author} {\bibfnamefont {U.}~\bibnamefont {Haisch}}, \ and\
 \bibinfo {author} {\bibfnamefont {J.}~\bibnamefont {Zupan}},\ }\href
 {\doibase 10.1007/JHEP11(2013)180} {\bibfield {journal} {\bibinfo {journal}
 {JHEP}\ }\textbf {\bibinfo {volume} {11}},\ \bibinfo {pages} {180} (\bibinfo
 {year} {2013})},\ \Eprint {http://arxiv.org/abs/1310.1385} {arXiv:1310.1385
 [hep-ph]}\BibitemShut {NoStop}%
\bibitem [{\citenamefont {Cirigliano}\ \emph
 {et~al.}(2016{\natexlab{a}})\citenamefont {Cirigliano}, \citenamefont
 {Dekens}, \citenamefont {de~Vries},\ and\ \citenamefont
 {Mereghetti}}]{Cirigliano:2016njn}%
 \BibitemOpen
 \bibfield {author} {\bibinfo {author} {\bibfnamefont {V.}~\bibnamefont
 {Cirigliano}}, \bibinfo {author} {\bibfnamefont {W.}~\bibnamefont {Dekens}},
 \bibinfo {author} {\bibfnamefont {J.}~\bibnamefont {de~Vries}}, \ and\
 \bibinfo {author} {\bibfnamefont {E.}~\bibnamefont {Mereghetti}},\ }\href
 {\doibase 10.1103/PhysRevD.94.016002} {\bibfield {journal} {\bibinfo
 {journal} {Phys. Rev. D}\ }\textbf {\bibinfo {volume} {94}},\ \bibinfo
 {pages} {016002} (\bibinfo {year} {2016}{\natexlab{a}})},\ \Eprint
 {http://arxiv.org/abs/1603.03049} {arXiv:1603.03049 [hep-ph]}\BibitemShut
 {NoStop}%
\bibitem [{\citenamefont {Cirigliano}\ \emph
 {et~al.}(2016{\natexlab{b}})\citenamefont {Cirigliano}, \citenamefont
 {Dekens}, \citenamefont {de~Vries},\ and\ \citenamefont
 {Mereghetti}}]{Cirigliano:2016nyn}%
 \BibitemOpen
 \bibfield {author} {\bibinfo {author} {\bibfnamefont {V.}~\bibnamefont
 {Cirigliano}}, \bibinfo {author} {\bibfnamefont {W.}~\bibnamefont {Dekens}},
 \bibinfo {author} {\bibfnamefont {J.}~\bibnamefont {de~Vries}}, \ and\
 \bibinfo {author} {\bibfnamefont {E.}~\bibnamefont {Mereghetti}},\ }\href
 {\doibase 10.1103/PhysRevD.94.034031} {\bibfield {journal} {\bibinfo
 {journal} {Phys. Rev. D}\ }\textbf {\bibinfo {volume} {94}},\ \bibinfo
 {pages} {034031} (\bibinfo {year} {2016}{\natexlab{b}})},\ \Eprint
 {http://arxiv.org/abs/1605.04311} {arXiv:1605.04311 [hep-ph]}\BibitemShut
 {NoStop}%
\bibitem [{\citenamefont {Buras}\ and\ \citenamefont
 {Jung}(2018)}]{Buras:2018gto}%
 \BibitemOpen
 \bibfield {author} {\bibinfo {author} {\bibfnamefont {A.~J.}\ \bibnamefont
 {Buras}}\ and\ \bibinfo {author} {\bibfnamefont {M.}~\bibnamefont {Jung}},\
 }\href {\doibase 10.1007/JHEP06(2018)067} {\bibfield {journal} {\bibinfo
 {journal} {JHEP}\ }\textbf {\bibinfo {volume} {06}},\ \bibinfo {pages} {067}
 (\bibinfo {year} {2018})},\ \Eprint {http://arxiv.org/abs/1804.05852}
 {arXiv:1804.05852 [hep-ph]}\BibitemShut {NoStop}%
\bibitem [{\citenamefont {Panico}\ \emph {et~al.}(2019)\citenamefont {Panico},
 \citenamefont {Pomarol},\ and\ \citenamefont {Riembau}}]{Panico:2018hal}%
 \BibitemOpen
 \bibfield {author} {\bibinfo {author} {\bibfnamefont {G.}~\bibnamefont
 {Panico}}, \bibinfo {author} {\bibfnamefont {A.}~\bibnamefont {Pomarol}}, \
 and\ \bibinfo {author} {\bibfnamefont {M.}~\bibnamefont {Riembau}},\ }\href
 {\doibase 10.1007/JHEP04(2019)090} {\bibfield {journal} {\bibinfo {journal}
 {JHEP}\ }\textbf {\bibinfo {volume} {04}},\ \bibinfo {pages} {090} (\bibinfo
 {year} {2019})},\ \Eprint {http://arxiv.org/abs/1810.09413} {arXiv:1810.09413
 [hep-ph]}\BibitemShut {NoStop}%
\bibitem [{\citenamefont {Brod}\ and\ \citenamefont
 {Stamou}(2021)}]{Brod:2018pli}%
 \BibitemOpen
 \bibfield {author} {\bibinfo {author} {\bibfnamefont {J.}~\bibnamefont
 {Brod}}\ and\ \bibinfo {author} {\bibfnamefont {E.}~\bibnamefont {Stamou}},\
 }\href {\doibase 10.1007/JHEP07(2021)080} {\bibfield {journal} {\bibinfo
 {journal} {JHEP}\ }\textbf {\bibinfo {volume} {07}},\ \bibinfo {pages} {080}
 (\bibinfo {year} {2021})},\ \Eprint {http://arxiv.org/abs/1810.12303}
 {arXiv:1810.12303 [hep-ph]}\BibitemShut {NoStop}%
\bibitem [{\citenamefont {Bauer}\ \emph {et~al.}(2021)\citenamefont {Bauer},
 \citenamefont {Neubert}, \citenamefont {Renner}, \citenamefont {Schnubel},\
 and\ \citenamefont {Thamm}}]{Bauer:2020jbp}%
 \BibitemOpen
 \bibfield {author} {\bibinfo {author} {\bibfnamefont {M.}~\bibnamefont
 {Bauer}}, \bibinfo {author} {\bibfnamefont {M.}~\bibnamefont {Neubert}},
 \bibinfo {author} {\bibfnamefont {S.}~\bibnamefont {Renner}}, \bibinfo
 {author} {\bibfnamefont {M.}~\bibnamefont {Schnubel}}, \ and\ \bibinfo
 {author} {\bibfnamefont {A.}~\bibnamefont {Thamm}},\ }\href {\doibase
 10.1007/JHEP04(2021)063} {\bibfield {journal} {\bibinfo {journal} {JHEP}\
 }\textbf {\bibinfo {volume} {04}},\ \bibinfo {pages} {063} (\bibinfo {year}
 {2021})},\ \Eprint {http://arxiv.org/abs/2012.12272} {arXiv:2012.12272
 [hep-ph]}\BibitemShut {NoStop}%
\bibitem [{\citenamefont {Ardu}\ and\ \citenamefont
 {Davidson}(2021)}]{Ardu:2021koz}%
 \BibitemOpen
 \bibfield {author} {\bibinfo {author} {\bibfnamefont {M.}~\bibnamefont
 {Ardu}}\ and\ \bibinfo {author} {\bibfnamefont {S.}~\bibnamefont
 {Davidson}},\ }\href {\doibase 10.1007/JHEP08(2021)002} {\bibfield {journal}
 {\bibinfo {journal} {JHEP}\ }\textbf {\bibinfo {volume} {08}},\ \bibinfo
 {pages} {002} (\bibinfo {year} {2021})},\ \Eprint
 {http://arxiv.org/abs/2103.07212} {arXiv:2103.07212 [hep-ph]}\BibitemShut
 {NoStop}%
 \bibitem [{\citenamefont {Allwicher}\ \emph {et~al.}(2023)\citenamefont
  {Allwicher}, \citenamefont {Isidori}, \citenamefont {Lizana}, \citenamefont
  {Selimovic},\ and\ \citenamefont {Stefanek}}]{Allwicher:2023aql}%
  \BibitemOpen
  \bibfield  {author} {\bibinfo {author} {\bibfnamefont {L.}~\bibnamefont
  {Allwicher}}, \bibinfo {author} {\bibfnamefont {G.}~\bibnamefont {Isidori}},
  \bibinfo {author} {\bibfnamefont {J.~M.}\ \bibnamefont {Lizana}}, \bibinfo
  {author} {\bibfnamefont {N.}~\bibnamefont {Selimovic}}, \ and\ \bibinfo
  {author} {\bibfnamefont {B.~A.}\ \bibnamefont {Stefanek}},\ }\href {\doibase
  10.1007/JHEP05(2023)179} {\bibfield  {journal} {\bibinfo  {journal} {JHEP}\
  }\textbf {\bibinfo {volume} {05}},\ \bibinfo {pages} {179} (\bibinfo {year}
  {2023})},\ \Eprint {http://arxiv.org/abs/2302.11584} {arXiv:2302.11584
  [hep-ph]}\BibitemShut {NoStop}%
  \BibitemOpen  
\bibitem [{\citenamefont {Brod}\ \emph {et~al.}(2024)\citenamefont {Brod},
 \citenamefont {Polonsky},\ and\ \citenamefont {Stamou}}]{Brod:2023wsh}%
 \bibfield {author} {\bibinfo {author} {\bibfnamefont {J.}~\bibnamefont
 {Brod}}, \bibinfo {author} {\bibfnamefont {Z.}~\bibnamefont {Polonsky}}, \
 and\ \bibinfo {author} {\bibfnamefont {E.}~\bibnamefont {Stamou}},\ }\href
 {\doibase 10.1007/JHEP06(2024)091} {\bibfield {journal} {\bibinfo {journal}
 {JHEP}\ }\textbf {\bibinfo {volume} {06}},\ \bibinfo {pages} {091} (\bibinfo
 {year} {2024})},\ \Eprint {http://arxiv.org/abs/2306.12478} {arXiv:2306.12478
 [hep-ph]}\BibitemShut {NoStop}%
\bibitem [{\citenamefont {Biek\"otter}\ \emph {et~al.}(2023)\citenamefont
 {Biek\"otter}, \citenamefont {Fuentes-Mart\'\i{}n}, \citenamefont {Galda},\
 and\ \citenamefont {Neubert}}]{Biekotter:2023mpd}%
 \BibitemOpen
 \bibfield {author} {\bibinfo {author} {\bibfnamefont {A.}~\bibnamefont
 {Biek\"otter}}, \bibinfo {author} {\bibfnamefont {J.}~\bibnamefont
 {Fuentes-Mart\'\i{}n}}, \bibinfo {author} {\bibfnamefont {A.~M.}\
 \bibnamefont {Galda}}, \ and\ \bibinfo {author} {\bibfnamefont
 {M.}~\bibnamefont {Neubert}},\ }\href {\doibase 10.1007/JHEP09(2023)120}
 {\bibfield {journal} {\bibinfo {journal} {JHEP}\ }\textbf {\bibinfo
 {volume} {09}},\ \bibinfo {pages} {120} (\bibinfo {year} {2023})},\ \Eprint
 {http://arxiv.org/abs/2307.10372} {arXiv:2307.10372 [hep-ph]}\BibitemShut
 {NoStop}%
\bibitem [{\citenamefont {Garosi}\ \emph {et~al.}(2023)\citenamefont {Garosi},
 \citenamefont {Marzocca}, \citenamefont {Rodriguez-Sanchez},\ and\
 \citenamefont {Stanzione}}]{Garosi:2023yxg}%
 \BibitemOpen
 \bibfield {author} {\bibinfo {author} {\bibfnamefont {F.}~\bibnamefont
 {Garosi}}, \bibinfo {author} {\bibfnamefont {D.}~\bibnamefont {Marzocca}},
 \bibinfo {author} {\bibfnamefont {A.}~\bibnamefont {Rodriguez-Sanchez}}, \
 and\ \bibinfo {author} {\bibfnamefont {A.}~\bibnamefont {Stanzione}},\ }\href
 {\doibase 10.1007/JHEP12(2023)129} {\bibfield {journal} {\bibinfo {journal}
 {JHEP}\ }\textbf {\bibinfo {volume} {12}},\ \bibinfo {pages} {129} (\bibinfo
 {year} {2023})},\ \Eprint {http://arxiv.org/abs/2310.00047} {arXiv:2310.00047
 [hep-ph]}\BibitemShut {NoStop}%
\bibitem [{\citenamefont {Allwicher}\ \emph {et~al.}(2024)\citenamefont
 {Allwicher}, \citenamefont {Cornella}, \citenamefont {Isidori},\ and\
 \citenamefont {Stefanek}}]{Allwicher:2023shc}%
 \BibitemOpen
 \bibfield {author} {\bibinfo {author} {\bibfnamefont {L.}~\bibnamefont
 {Allwicher}}, \bibinfo {author} {\bibfnamefont {C.}~\bibnamefont {Cornella}},
 \bibinfo {author} {\bibfnamefont {G.}~\bibnamefont {Isidori}}, \ and\
 \bibinfo {author} {\bibfnamefont {B.~A.}\ \bibnamefont {Stefanek}},\ }\href
 {\doibase 10.1007/JHEP03(2024)049} {\bibfield {journal} {\bibinfo {journal}
 {JHEP}\ }\textbf {\bibinfo {volume} {03}},\ \bibinfo {pages} {049} (\bibinfo
 {year} {2024})},\ \Eprint {http://arxiv.org/abs/2311.00020} {arXiv:2311.00020
 [hep-ph]}\BibitemShut {NoStop}%
\bibitem [{\citenamefont {Stefanek}(2024)}]{Stefanek:2024kds}%
 \BibitemOpen
 \bibfield {author} {\bibinfo {author} {\bibfnamefont {B.~A.}\ \bibnamefont
 {Stefanek}},\ }\href {\doibase 10.1007/JHEP09(2024)103} {\bibfield {journal}
 {\bibinfo {journal} {JHEP}\ }\textbf {\bibinfo {volume} {09}},\ \bibinfo
 {pages} {103} (\bibinfo {year} {2024})},\ \Eprint
 {http://arxiv.org/abs/2407.09593} {arXiv:2407.09593 [hep-ph]}\BibitemShut
 {NoStop}%
\bibitem [{\citenamefont {Fuentes-Martin}\ \emph {et~al.}(2021)\citenamefont
 {Fuentes-Martin}, \citenamefont {Ruiz-Femenia}, \citenamefont {Vicente},\
 and\ \citenamefont {Virto}}]{Fuentes-Martin:2020zaz}%
 \BibitemOpen
 \bibfield {author} {\bibinfo {author} {\bibfnamefont {J.}~\bibnamefont
 {Fuentes-Martin}}, \bibinfo {author} {\bibfnamefont {P.}~\bibnamefont
 {Ruiz-Femenia}}, \bibinfo {author} {\bibfnamefont {A.}~\bibnamefont
 {Vicente}}, \ and\ \bibinfo {author} {\bibfnamefont {J.}~\bibnamefont
 {Virto}},\ }\href {\doibase 10.1140/epjc/s10052-020-08778-y} {\bibfield
 {journal} {\bibinfo {journal} {Eur. Phys. J. C}\ }\textbf {\bibinfo {volume}
 {81}},\ \bibinfo {pages} {167} (\bibinfo {year} {2021})},\ \Eprint
 {http://arxiv.org/abs/2010.16341} {arXiv:2010.16341 [hep-ph]}\BibitemShut
 {NoStop}%
\bibitem [{\citenamefont {Navas}\ \emph {et~al.}(2024)\citenamefont {Navas}
 \emph {et~al.}}]{ParticleDataGroup:2024cfk}%
 \BibitemOpen
 \bibfield {author} {\bibinfo {author} {\bibfnamefont {S.}~\bibnamefont
 {Navas}} \emph {et~al.} (\bibinfo {collaboration} {Particle Data Group}),\
 }\href {\doibase 10.1103/PhysRevD.110.030001} {\bibfield {journal} {\bibinfo
 {journal} {Phys. Rev. D}\ }\textbf {\bibinfo {volume} {110}},\ \bibinfo
 {pages} {030001} (\bibinfo {year} {2024})}\BibitemShut {NoStop}%
\bibitem [{\citenamefont {Aad}\ \emph {et~al.}(2022)\citenamefont {Aad} \emph
 {et~al.}}]{ATLAS:2022vkf}%
 \BibitemOpen
 \bibfield {author} {\bibinfo {author} {\bibfnamefont {G.}~\bibnamefont
 {Aad}} \emph {et~al.} (\bibinfo {collaboration} {ATLAS}),\ }\href {\doibase
 10.1038/s41586-022-04893-w} {\bibfield {journal} {\bibinfo {journal}
 {Nature}\ }\textbf {\bibinfo {volume} {607}},\ \bibinfo {pages} {52}
 (\bibinfo {year} {2022})},\ \bibinfo {note} {[Erratum: Nature {\bf 612}, E24
 (2022)]},\ \Eprint {http://arxiv.org/abs/2207.00092} {arXiv:2207.00092
 [hep-ex]}\BibitemShut {NoStop}%
\bibitem [{\citenamefont {Maitre}(2006)}]{Maitre:2005uu}%
 \BibitemOpen
 \bibfield {author} {\bibinfo {author} {\bibfnamefont {D.}~\bibnamefont
 {Maitre}},\ }\href {\doibase 10.1016/j.cpc.2005.10.008} {\bibfield {journal}
 {\bibinfo {journal} {Comput. Phys. Commun.}\ }\textbf {\bibinfo {volume}
 {174}},\ \bibinfo {pages} {222} (\bibinfo {year} {2006})},\ \Eprint
 {http://arxiv.org/abs/hep-ph/0507152} {arXiv:hep-ph/0507152}\BibitemShut
 {NoStop}%
\bibitem [{\citenamefont {Cheng}\ \emph {et~al.}(1974)\citenamefont {Cheng},
 \citenamefont {Eichten},\ and\ \citenamefont {Li}}]{Cheng:1973nv}%
 \BibitemOpen
 \bibfield {author} {\bibinfo {author} {\bibfnamefont {T.~P.}\ \bibnamefont
 {Cheng}}, \bibinfo {author} {\bibfnamefont {E.}~\bibnamefont {Eichten}}, \
 and\ \bibinfo {author} {\bibfnamefont {L.-F.}\ \bibnamefont {Li}},\ }\href
 {\doibase 10.1103/PhysRevD.9.2259} {\bibfield {journal} {\bibinfo {journal}
 {Phys. Rev. D}\ }\textbf {\bibinfo {volume} {9}},\ \bibinfo {pages} {2259}
 (\bibinfo {year} {1974})}\BibitemShut {NoStop}%
\bibitem [{\citenamefont {Machacek}\ and\ \citenamefont
 {Vaughn}(1984)}]{Machacek:1983fi}%
 \BibitemOpen
 \bibfield {author} {\bibinfo {author} {\bibfnamefont {M.~E.}\ \bibnamefont
 {Machacek}}\ and\ \bibinfo {author} {\bibfnamefont {M.~T.}\ \bibnamefont
 {Vaughn}},\ }\href {\doibase 10.1016/0550-3213(84)90533-9} {\bibfield
 {journal} {\bibinfo {journal} {Nucl. Phys. B}\ }\textbf {\bibinfo {volume}
 {236}},\ \bibinfo {pages} {221} (\bibinfo {year} {1984})}\BibitemShut
 {NoStop}%
\bibitem [{\citenamefont {Pendleton}\ and\ \citenamefont
 {Ross}(1981)}]{Pendleton:1980as}%
 \BibitemOpen
 \bibfield {author} {\bibinfo {author} {\bibfnamefont {B.}~\bibnamefont
 {Pendleton}}\ and\ \bibinfo {author} {\bibfnamefont {G.~G.}\ \bibnamefont
 {Ross}},\ }\href {\doibase 10.1016/0370-2693(81)90017-4} {\bibfield
 {journal} {\bibinfo {journal} {Phys. Lett. B}\ }\textbf {\bibinfo {volume}
 {98}},\ \bibinfo {pages} {291} (\bibinfo {year} {1981})}\BibitemShut
 {NoStop}%
\bibitem [{\citenamefont {Hill}(1981)}]{Hill:1980sq}%
 \BibitemOpen
 \bibfield {author} {\bibinfo {author} {\bibfnamefont {C.~T.}\ \bibnamefont
 {Hill}},\ }\href {\doibase 10.1103/PhysRevD.24.691} {\bibfield {journal}
 {\bibinfo {journal} {Phys. Rev. D}\ }\textbf {\bibinfo {volume} {24}},\
 \bibinfo {pages} {691} (\bibinfo {year} {1981})}\BibitemShut {NoStop}%
\end{thebibliography}

%\end{document}

%merlin.mbs apsrev4-1.bst 2010-07-25 4.21a (PWD, AO, DPC) hacked
%Control: key (0)
%Control: author (72) initials jnrlst
%Control: editor formatted (1) identically to author
%Control: production of article title (-1) disabled
%Control: page (0) single
%Control: year (1) truncated
%Control: production of eprint (0) enabled
%

\end{document}